\documentclass[twocolumn,preprintnumbers,prl,aps]{revtex4-1}
\usepackage{graphicx,amsmath,bbm,xcolor,float}
\usepackage[breaklinks=true]{hyperref}
\usepackage{breakurl}
\graphicspath{ {fig/} }

\allowdisplaybreaks[4]

\definecolor{darkblue}{rgb}{0.0,0.0,0.7}
\newcommand{\CS}{\textcolor{darkblue}}

\newcommand{\Graph}[2]{\vcenter{\hbox{\includegraphics[scale=#1]{#2}}}}
\newcommand{\ff}{\bar{\mathcal{F}}}
\newcommand{\bigO}[1]{\mathcal{O}\left(#1\right)}
\newcommand{\HyperInt}{\texttt{HyperInt}}
\newcommand{\Finred}{\texttt{Finred}}

\newcounter{qCS} 
\newcounter{gCS} 
\newcommand{\qFFterm}[2]{%
    \stepcounter{qCS}%
    \global\expandafter\edef\csname qCS-#2\endcsname{\theqCS}%
    \CS{#1}\,\qFF{#2}
}
\newcommand{\gFFterm}[2]{%
    \stepcounter{gCS}%
    \global\expandafter\edef\csname gCS-#2\endcsname{\thegCS}%
    \CS{#1}\,\gFF{#2}
}
\newcommand{\qFF}[1]{%
    c^q_{\csname qCS-#1\endcsname}(\epsilon)
}
\newcommand{\gFF}[1]{%
    c^g_{\csname gCS-#1\endcsname}(\epsilon)
}

\begin{document}

\preprint{MSUHEP-20-002}

\title{Cusp and collinear anomalous dimensions in four-loop QCD from form factors}

\author{Andreas von Manteuffel,$^{\scriptstyle 1}$ Erik Panzer,$^{\scriptstyle 2}$ and Robert M. Schabinger$\,^{\scriptstyle 1}$} 

\affiliation{$^{\scriptstyle 1}$Department of Physics and Astronomy, Michigan State University, East Lansing, Michigan 48824, USA\\
$^{\scriptstyle 2}$All Souls College, University of Oxford, OX1 4AL, Oxford, UK}

\begin{abstract}
\noindent
We calculate the complete quark and gluon cusp anomalous dimensions in four-loop massless QCD analytically from first principles.
In addition, we determine the complete matter dependence of the quark and gluon collinear anomalous dimensions.
Our approach is to Laurent expand four-loop quark and gluon form factors in the parameter of dimensional regularization.
We employ finite field and syzygy techniques to reduce the relevant Feynman integrals to a basis of finite integrals,
and subsequently evaluate the basis integrals directly from their standard parametric representations.
\end{abstract}

\maketitle
\section{Introduction}

\begin{figure*}
\begin{align*}
&\quad \Graph{.175}{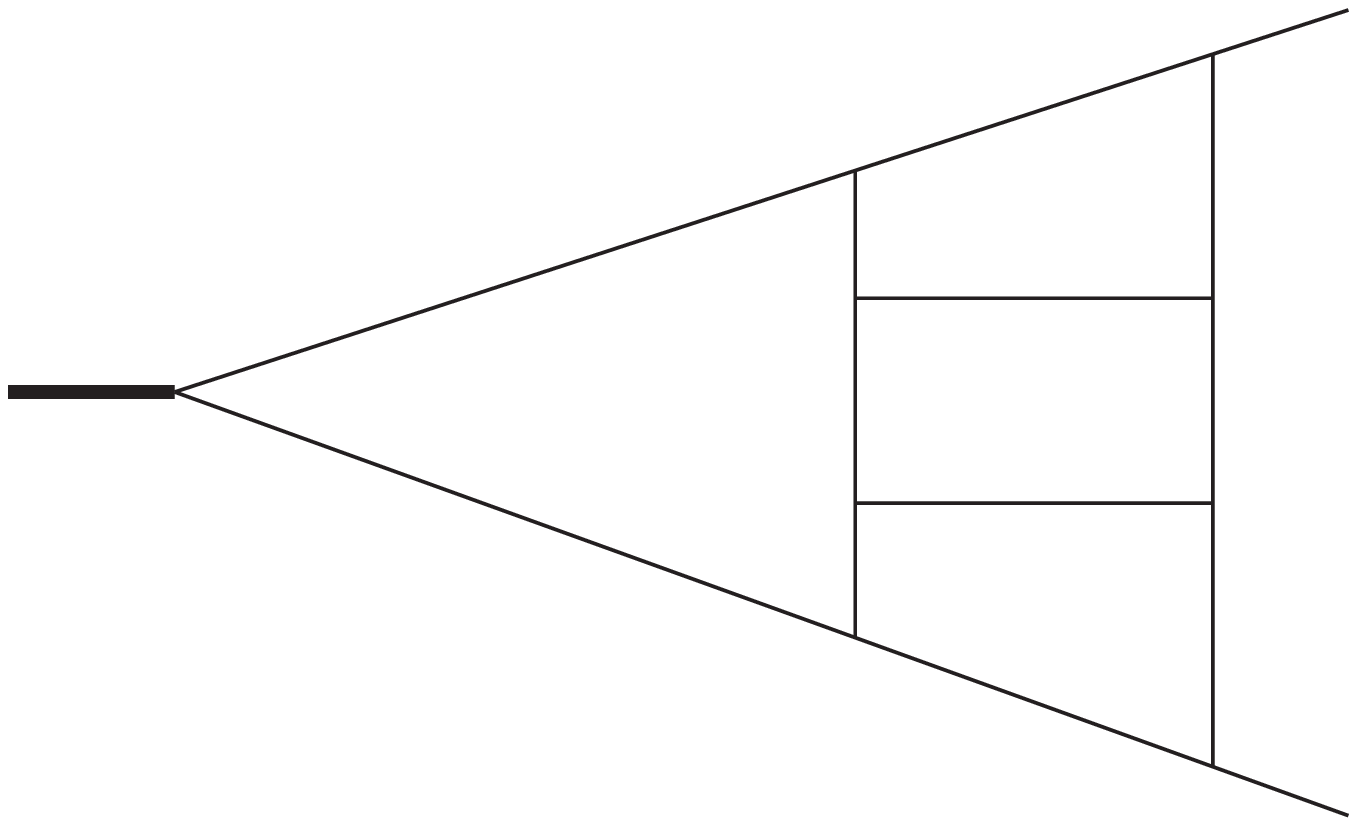} \qquad \Graph{.175}{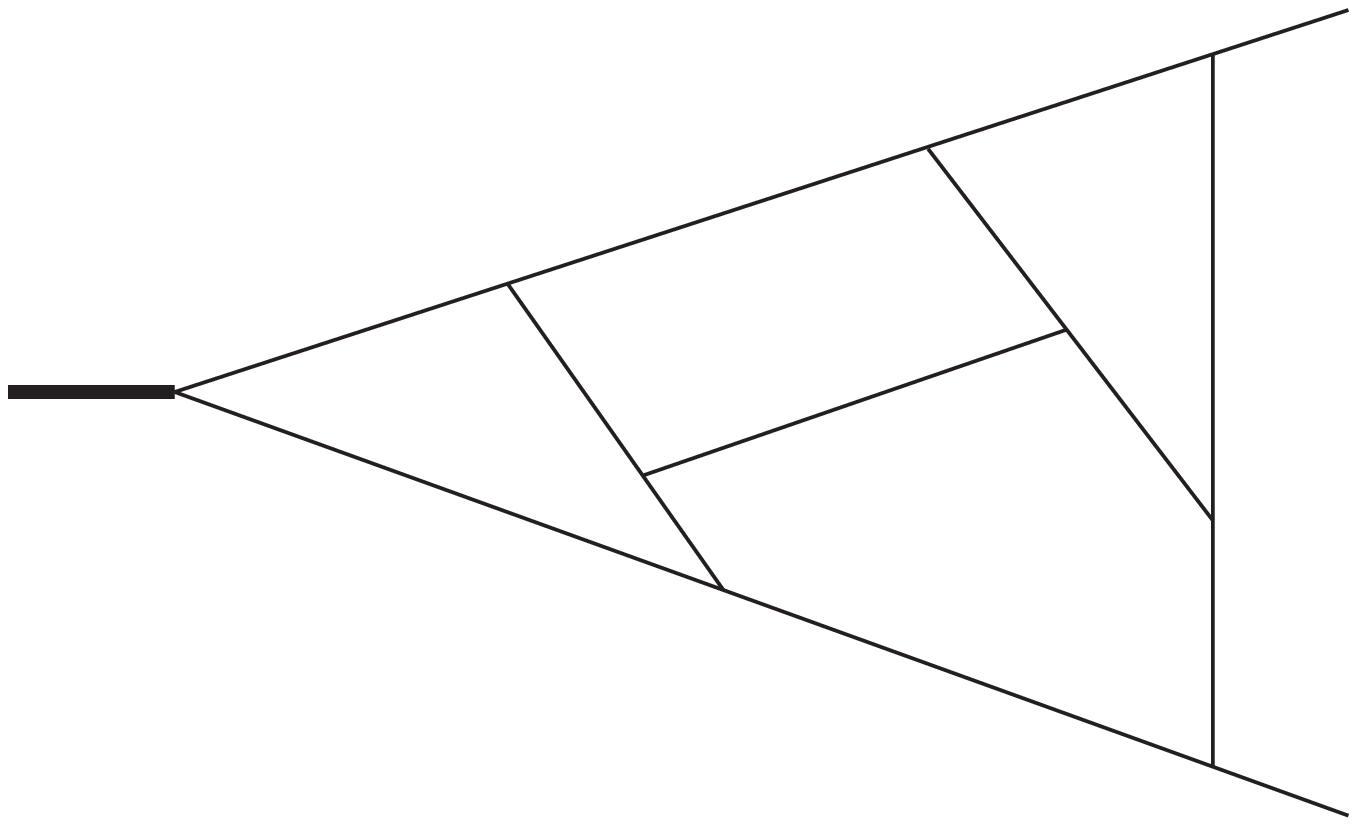}  \qquad \Graph{.175}{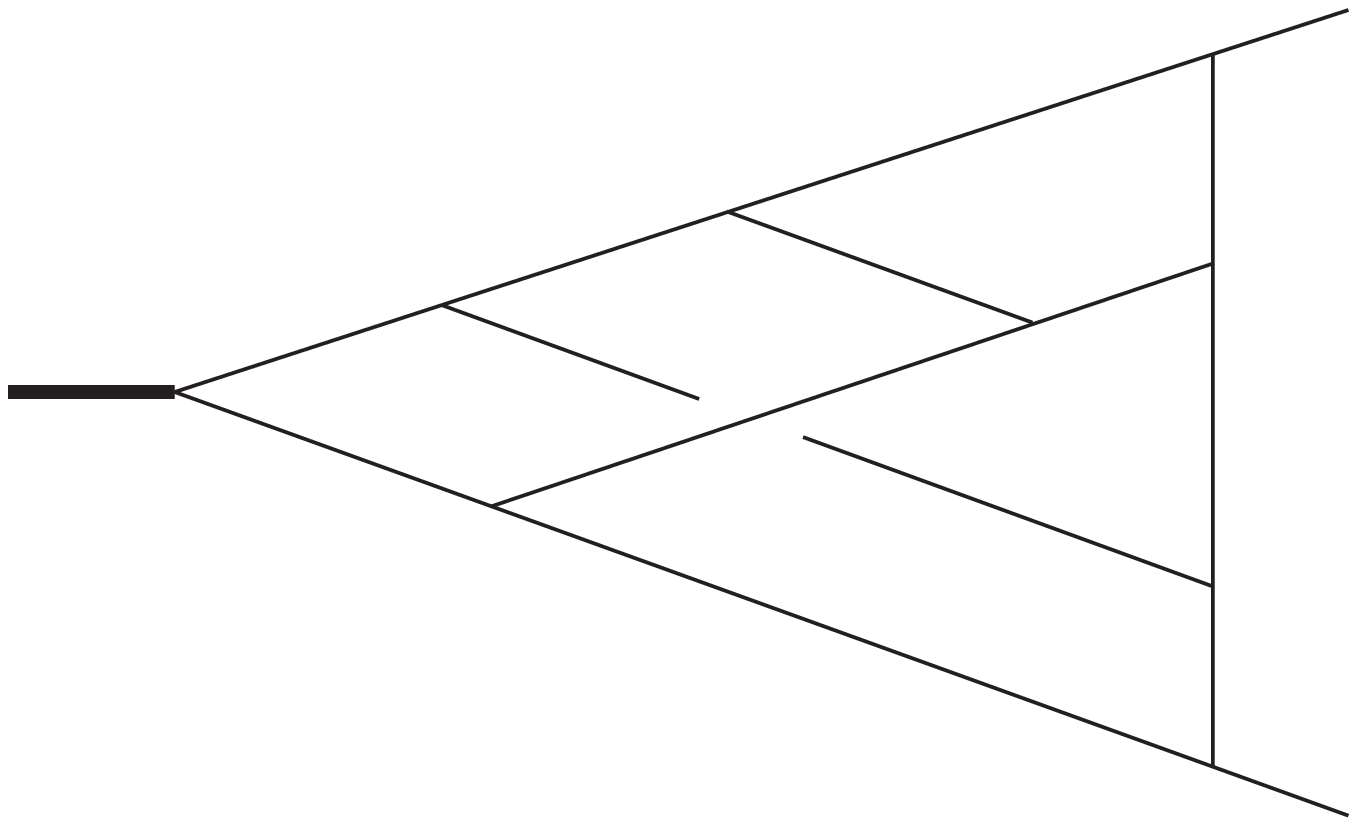} \qquad \Graph{.175}{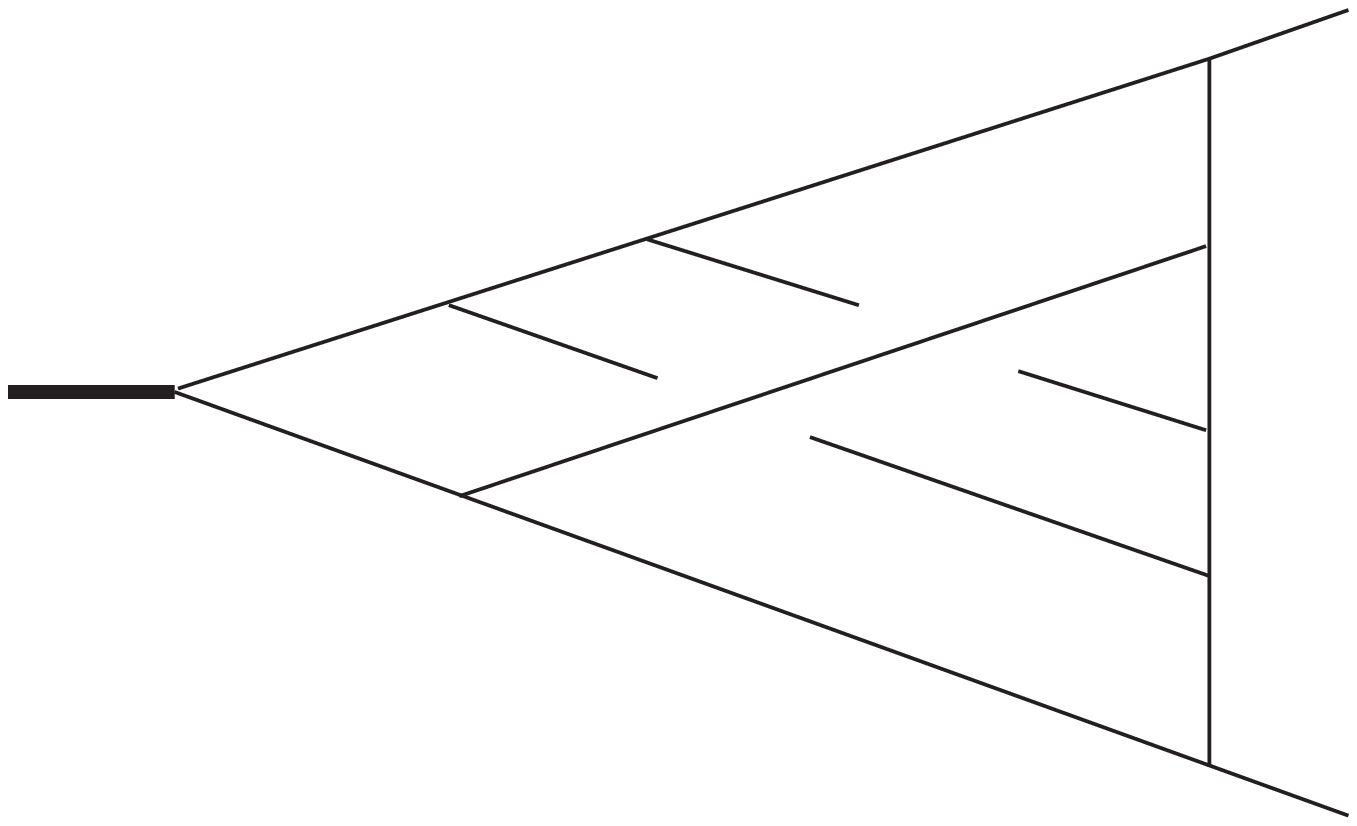} \qquad \Graph{.175}{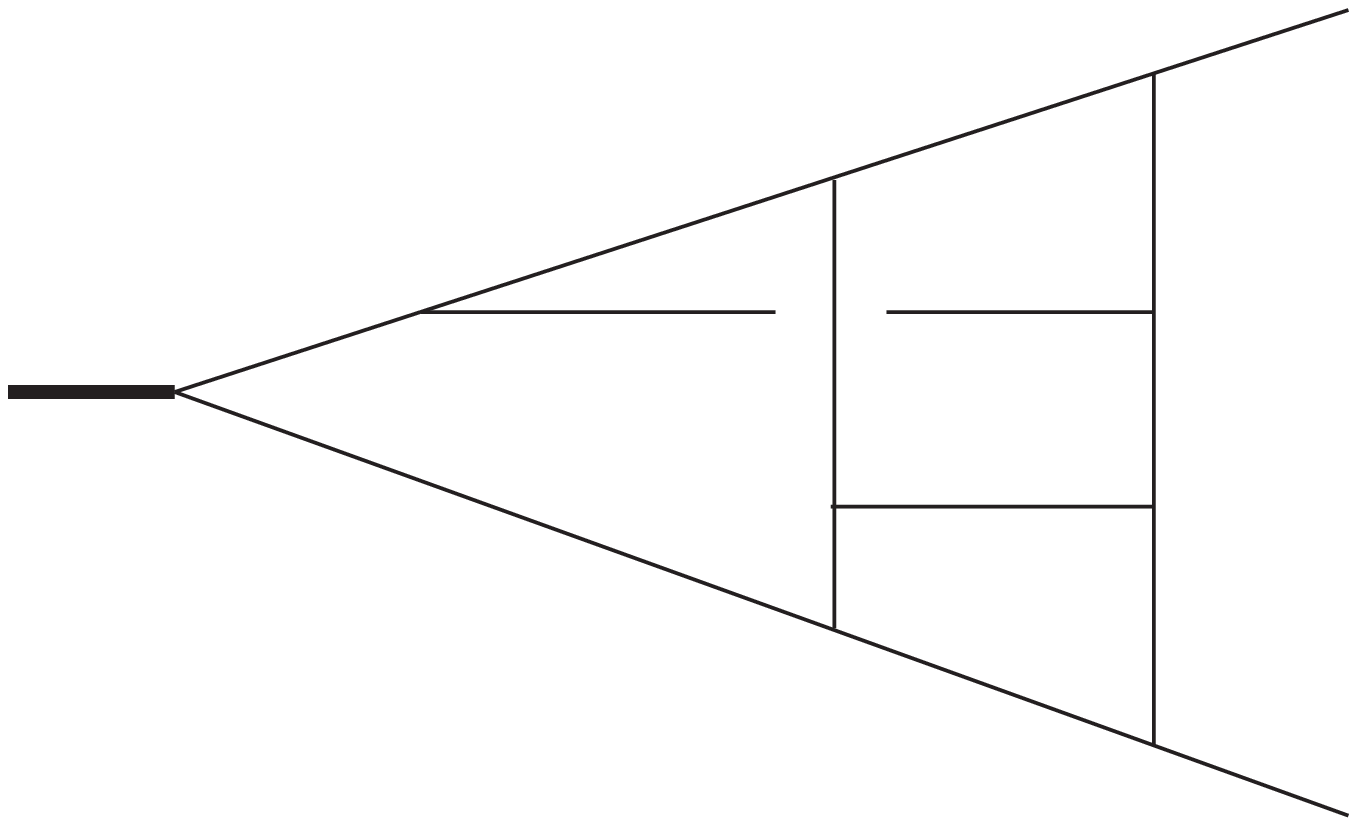} \\
&\quad \Graph{.175}{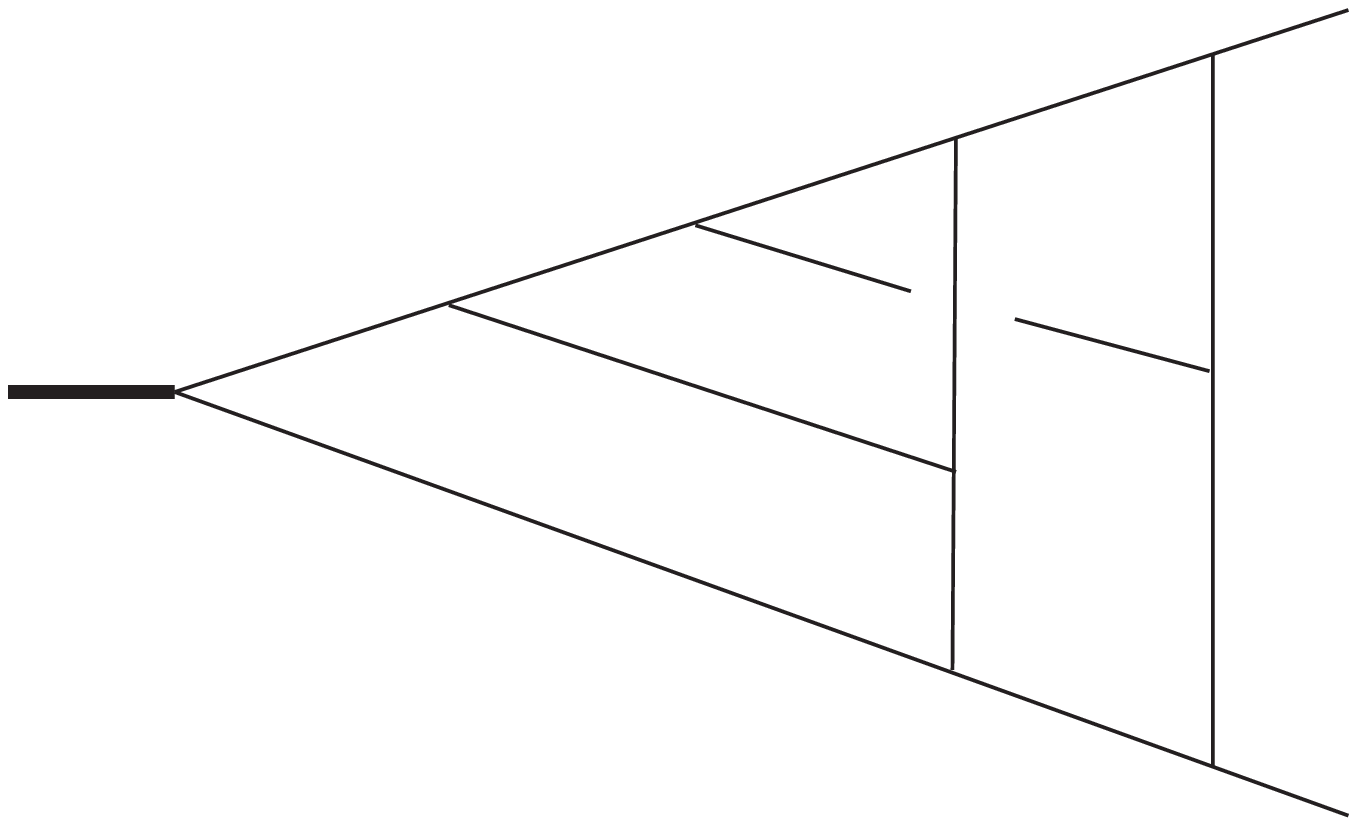} \qquad \Graph{.175}{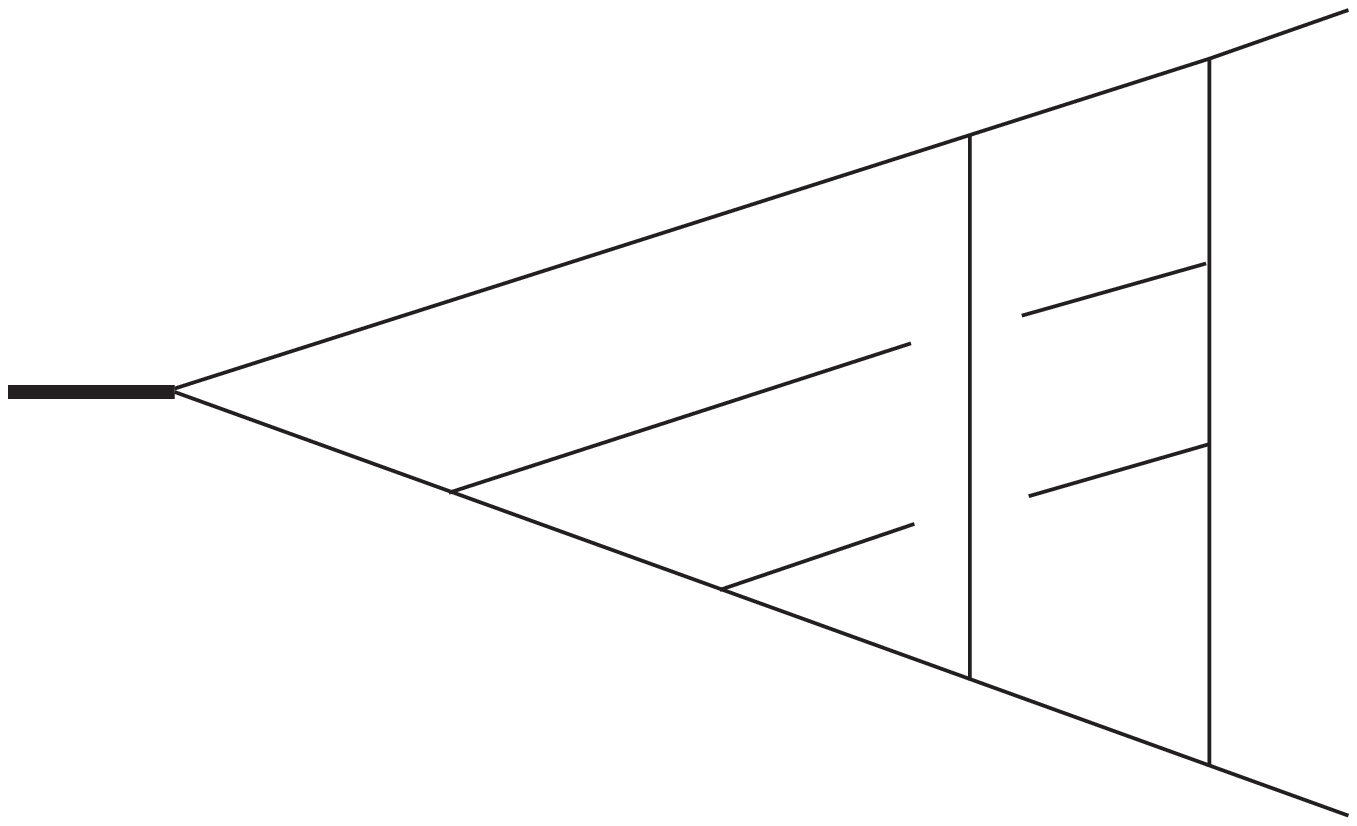} \qquad \Graph{.175}{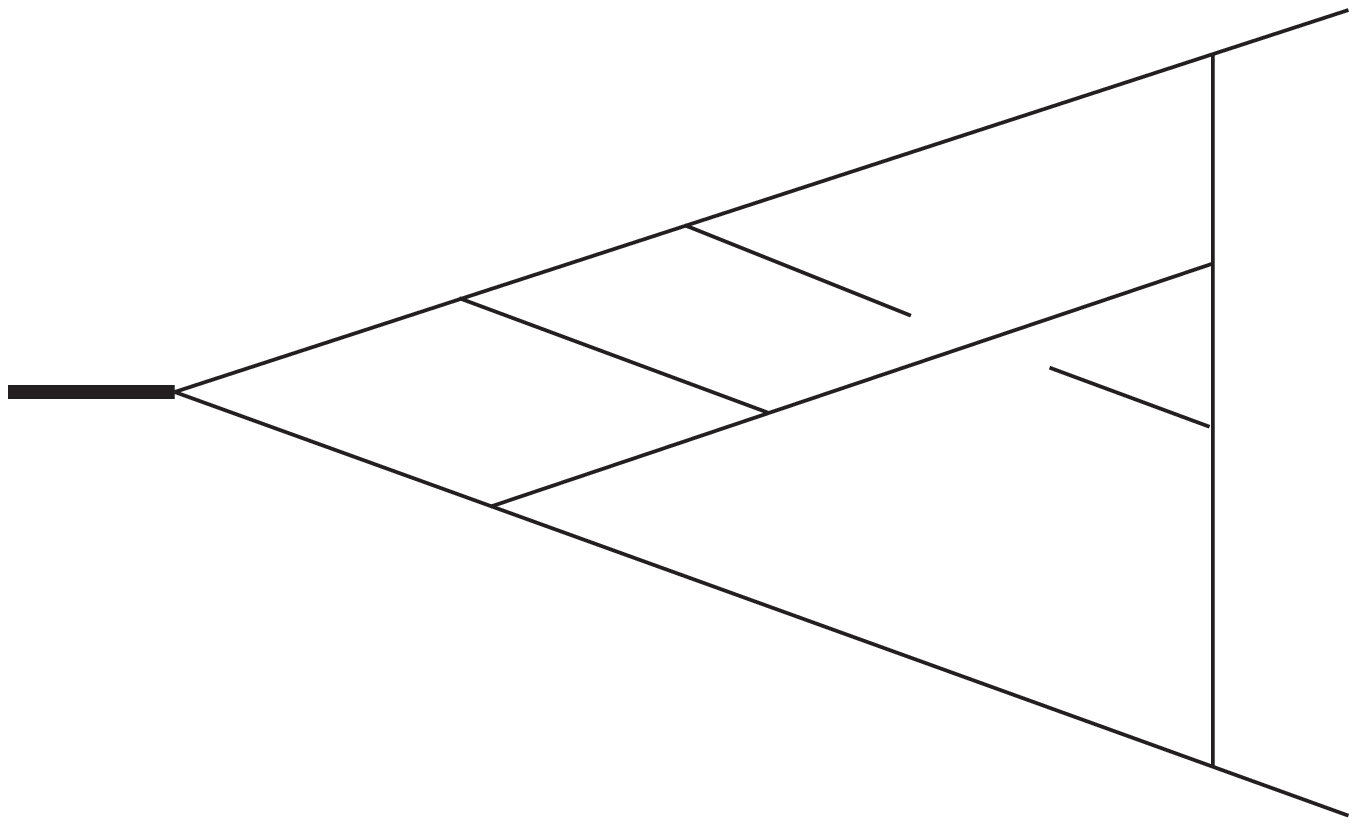} 
\qquad \!\!\fbox{$\Graph{.175}{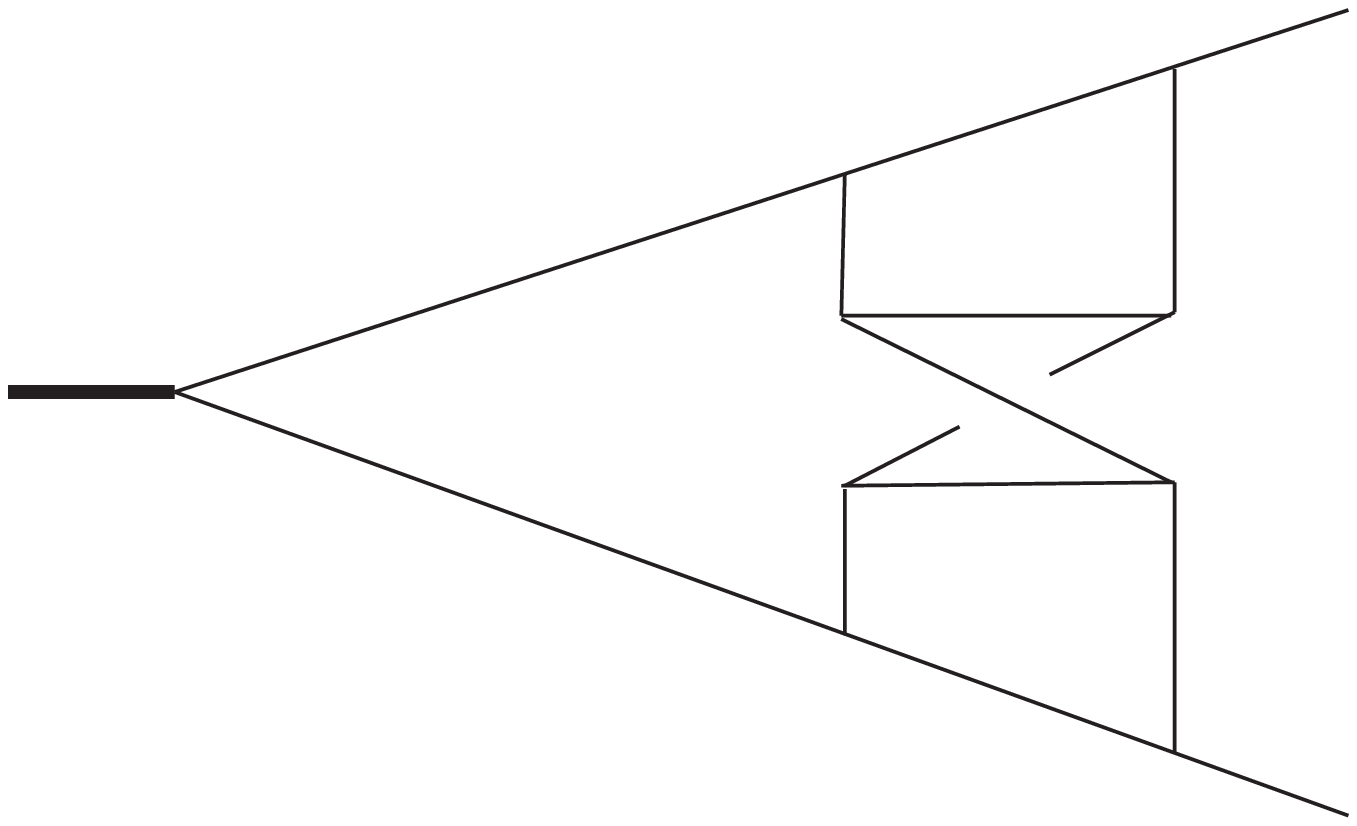} \qquad \Graph{.175}{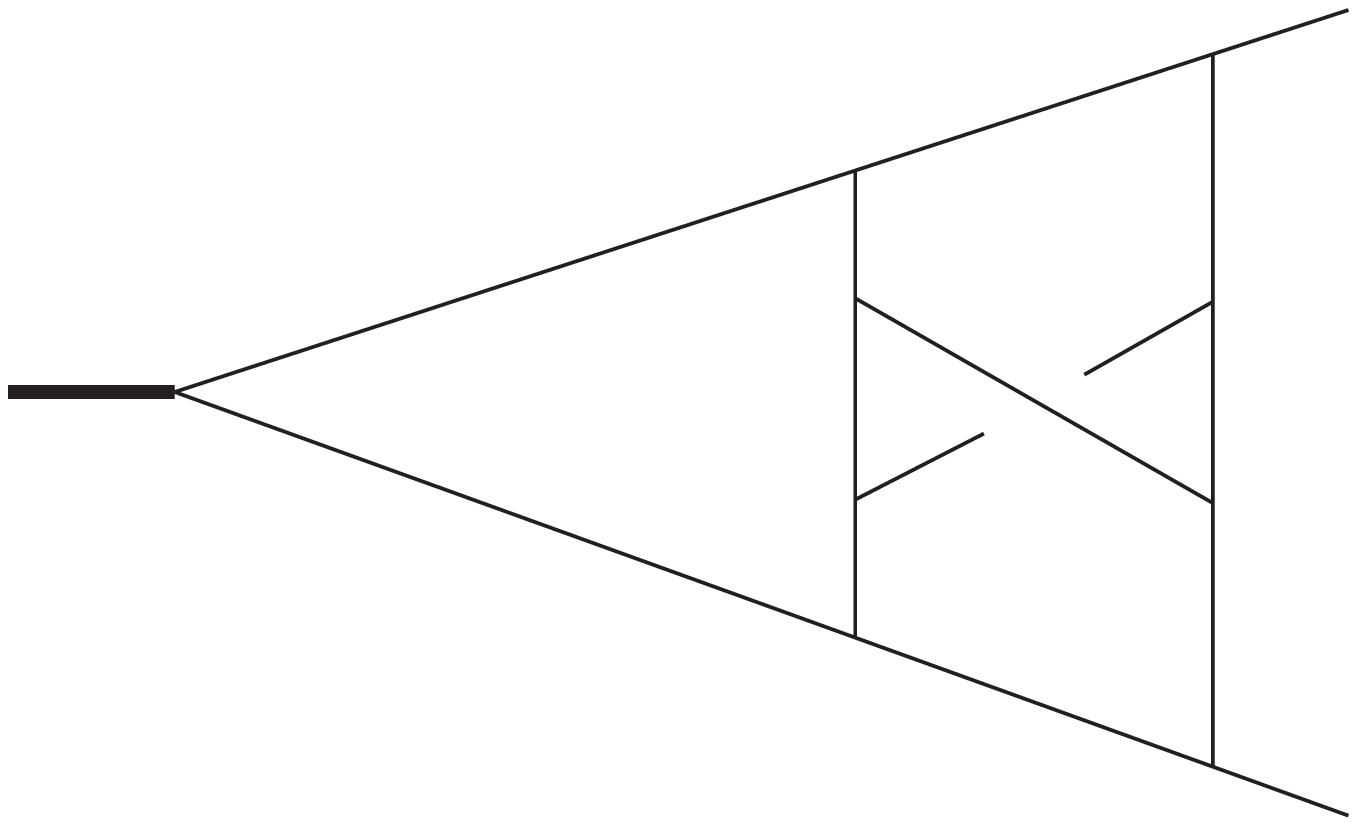}$}
\Leftarrow {\parbox{19mm}{top-level topologies contributing to $\Gamma_4^q$ and $\Gamma_4^g$}}
\\
&\quad \Graph{.175}{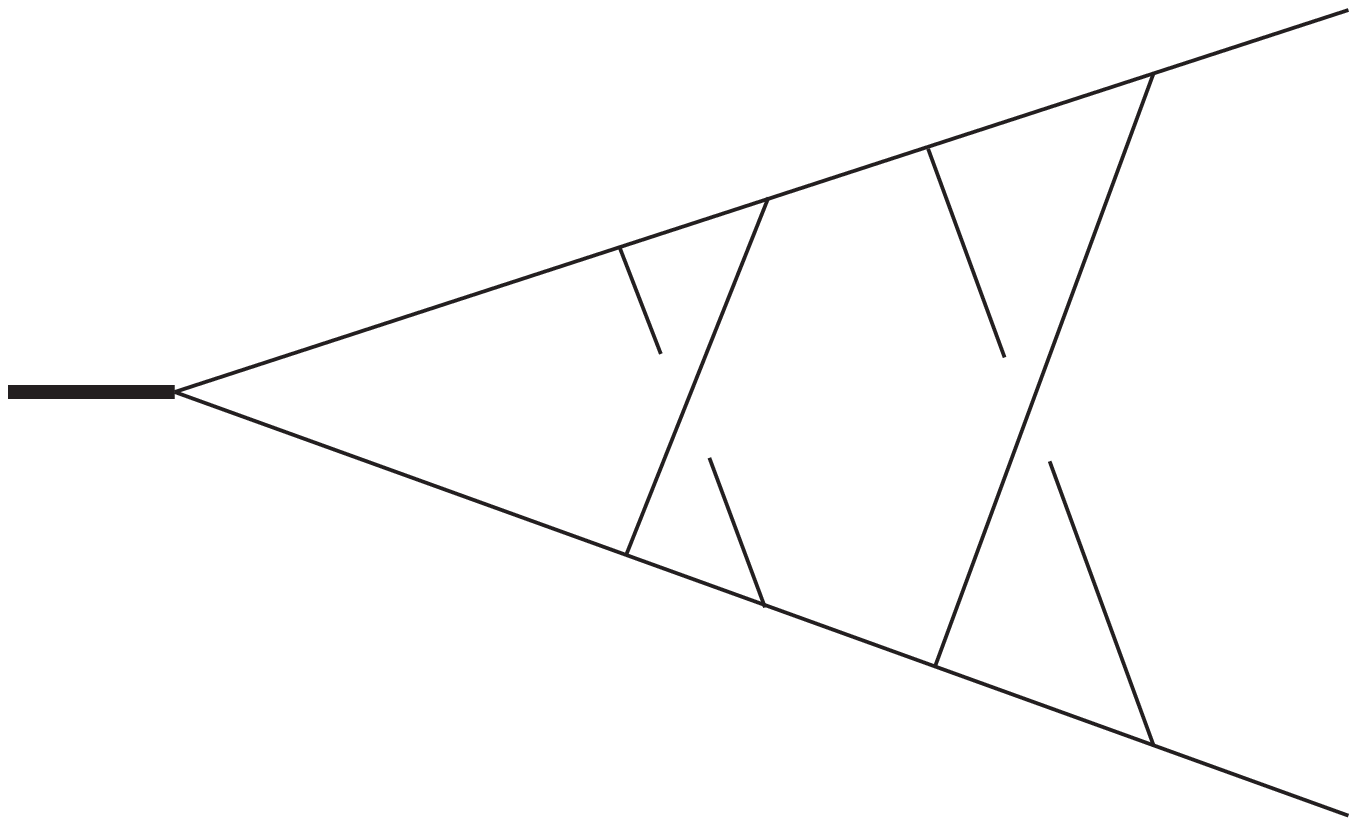} \qquad \Graph{.175}{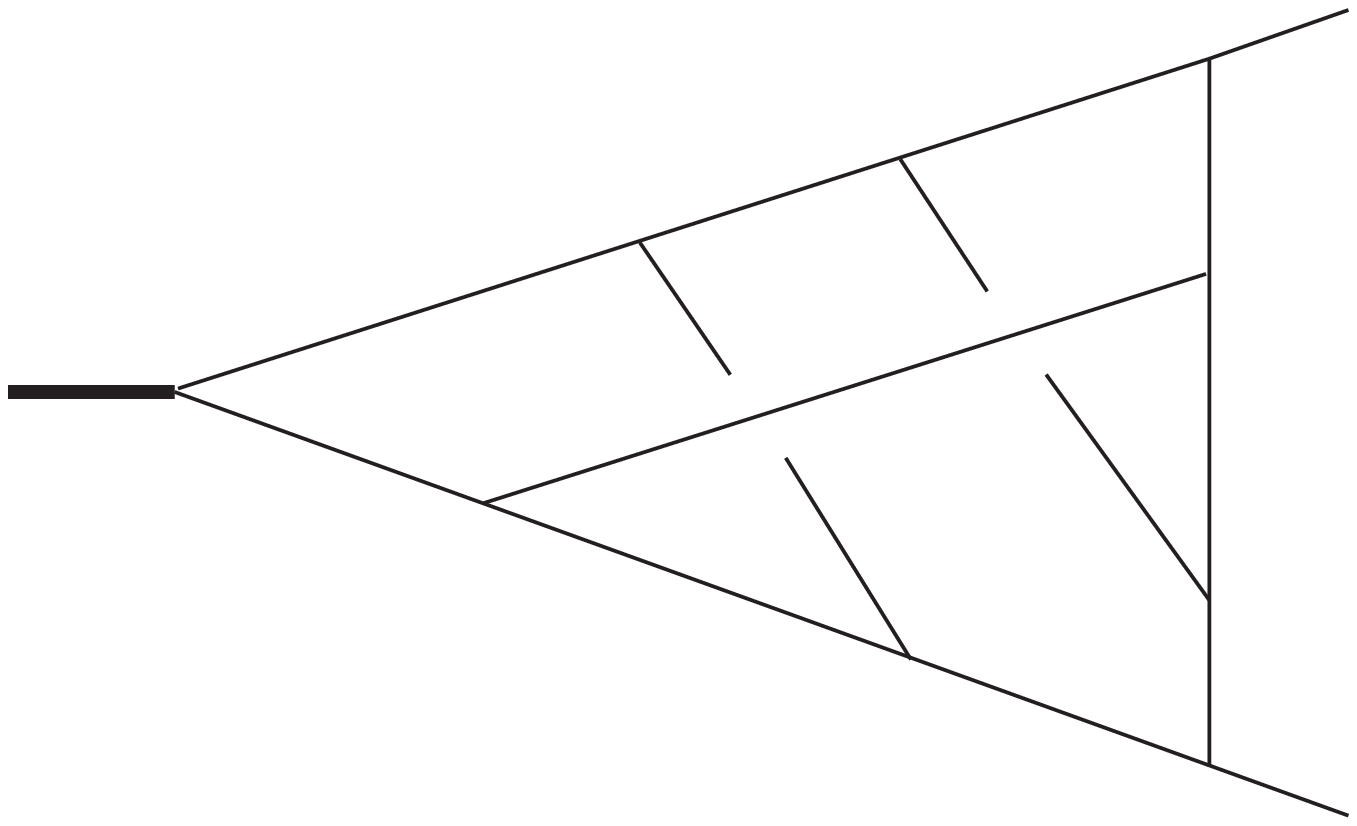} \qquad \Graph{.175}{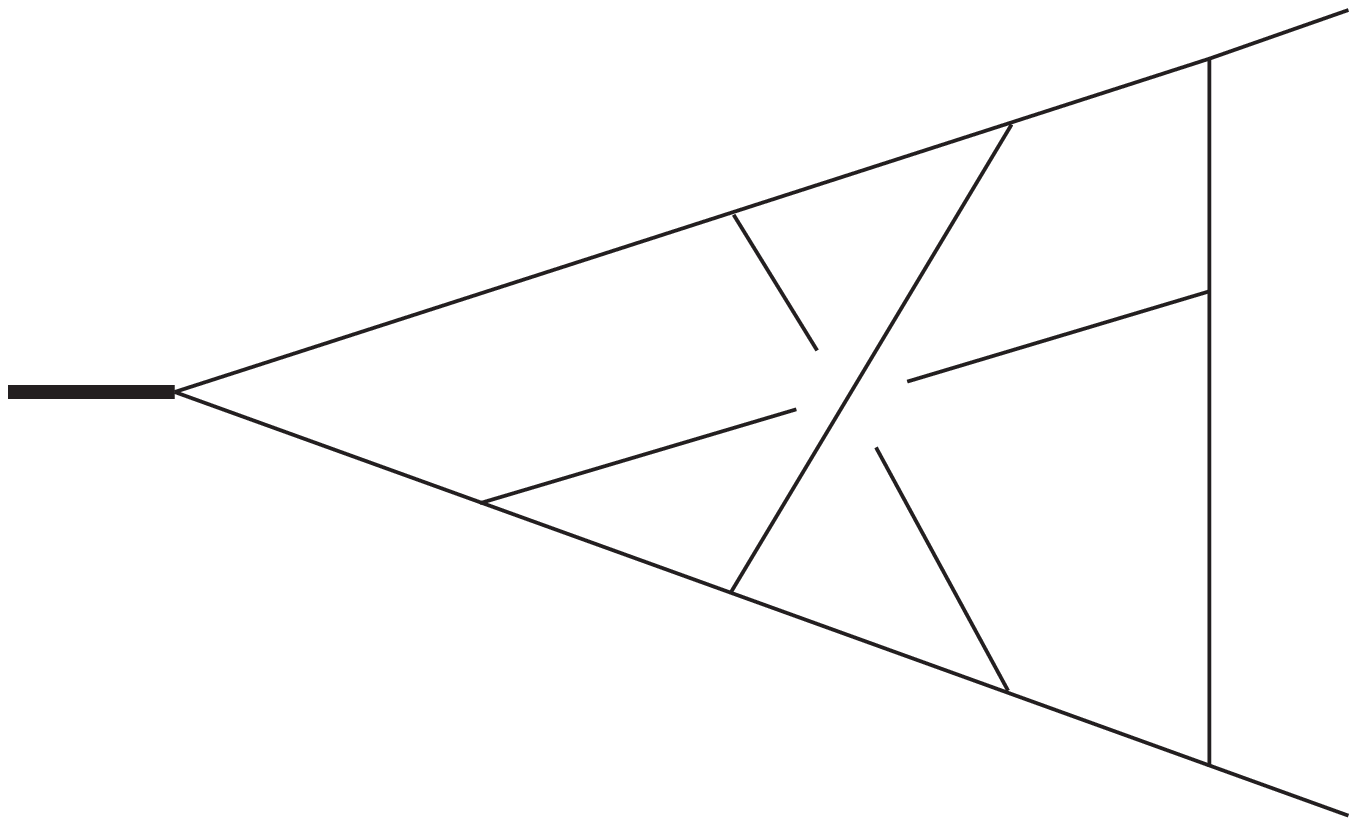} \qquad \Graph{.175}{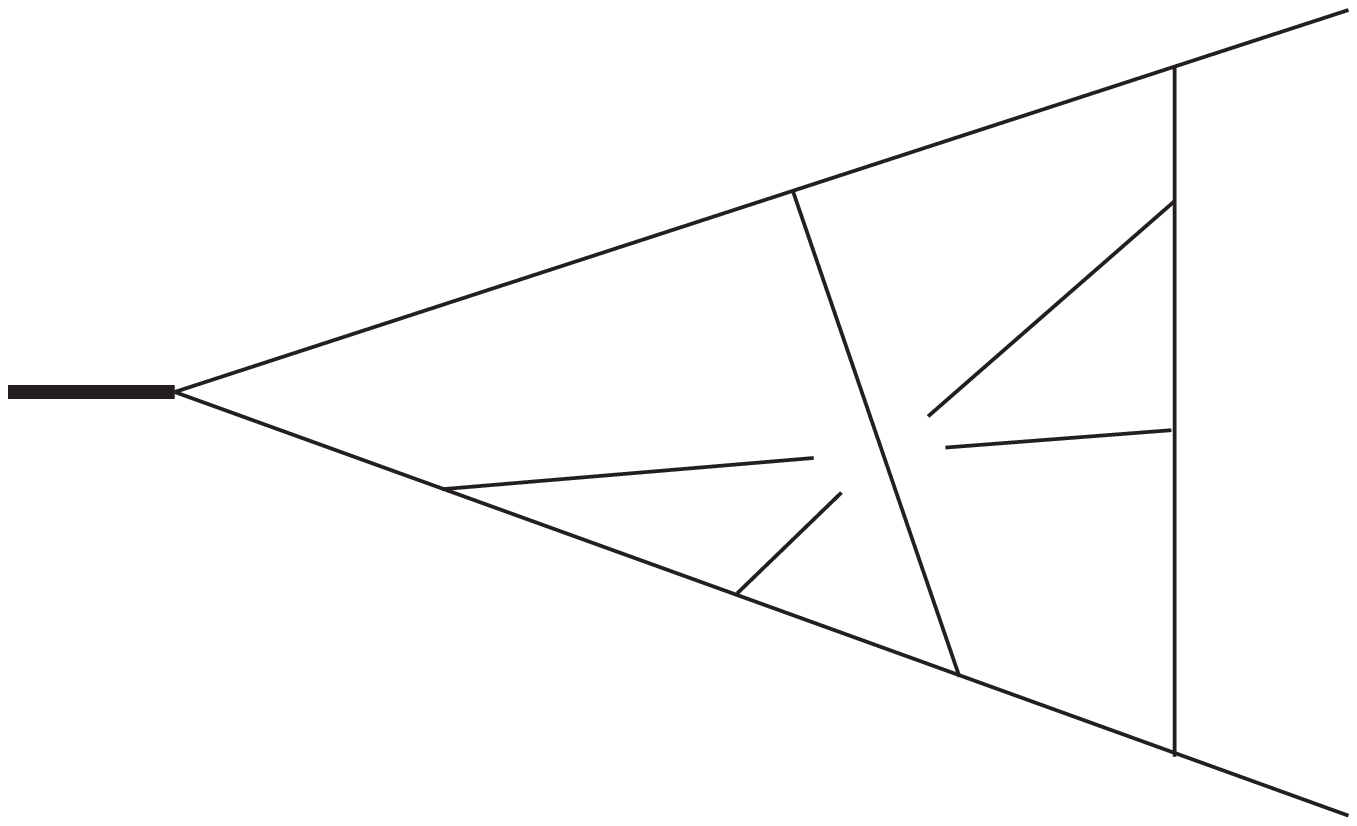} \qquad \Graph{.175}{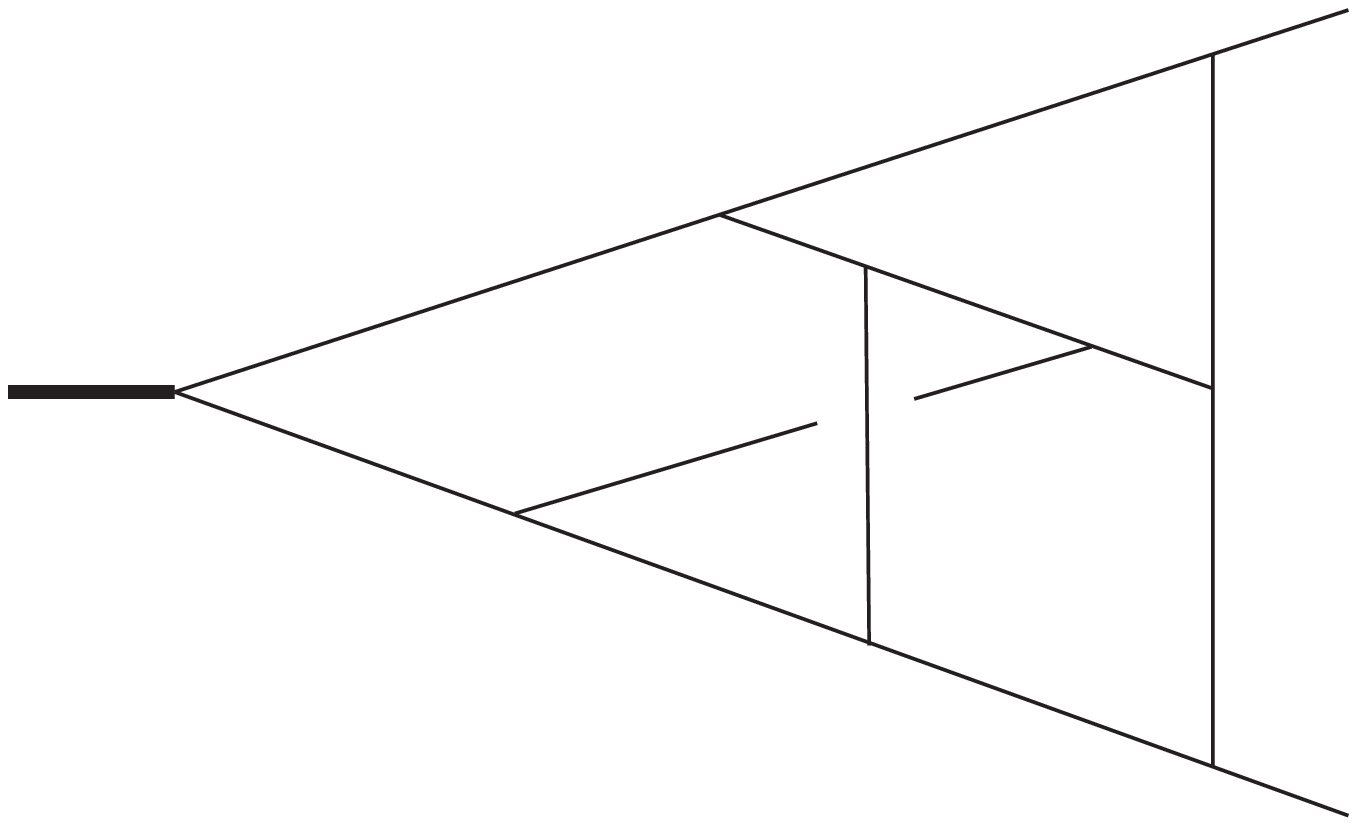}\\
&\quad \Graph{.175}{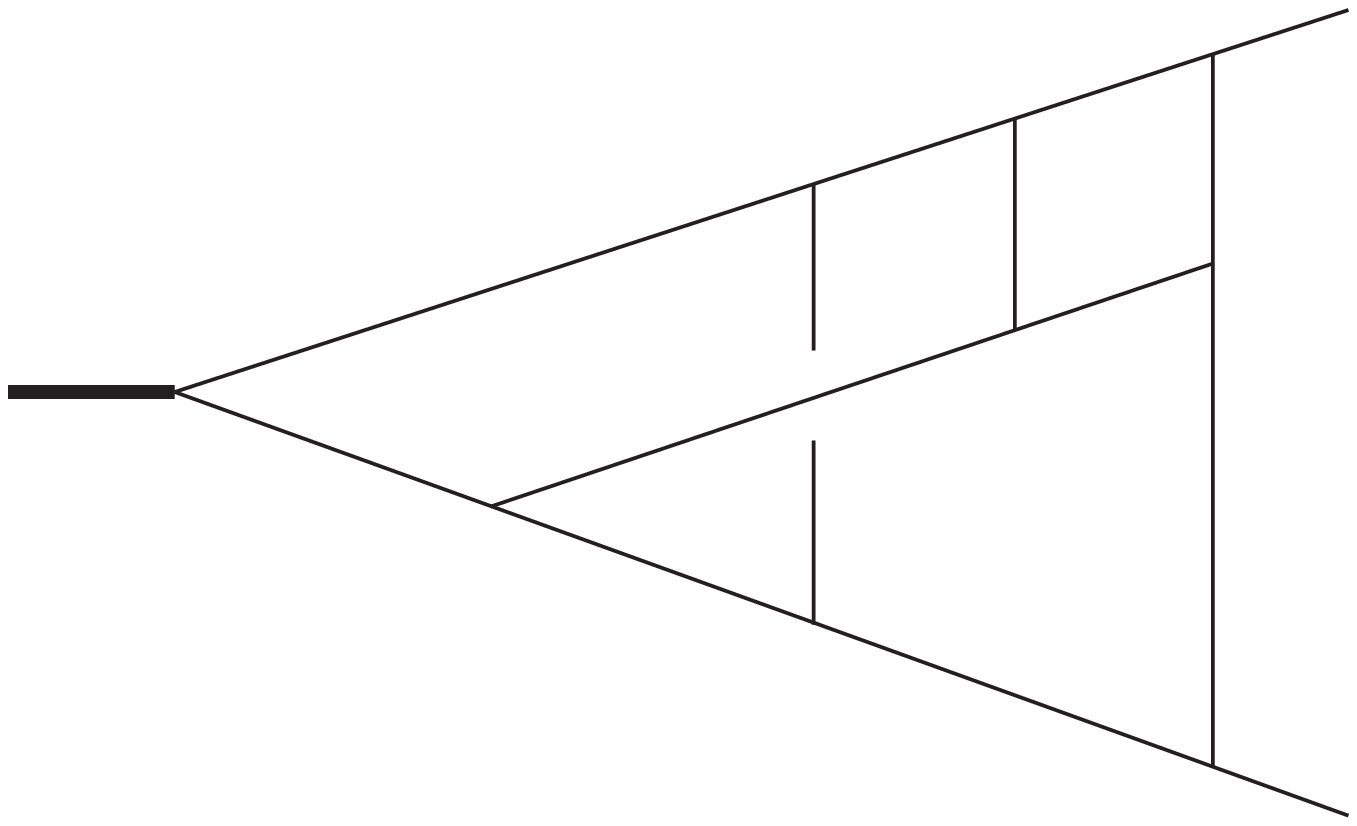} \qquad \Graph{.175}{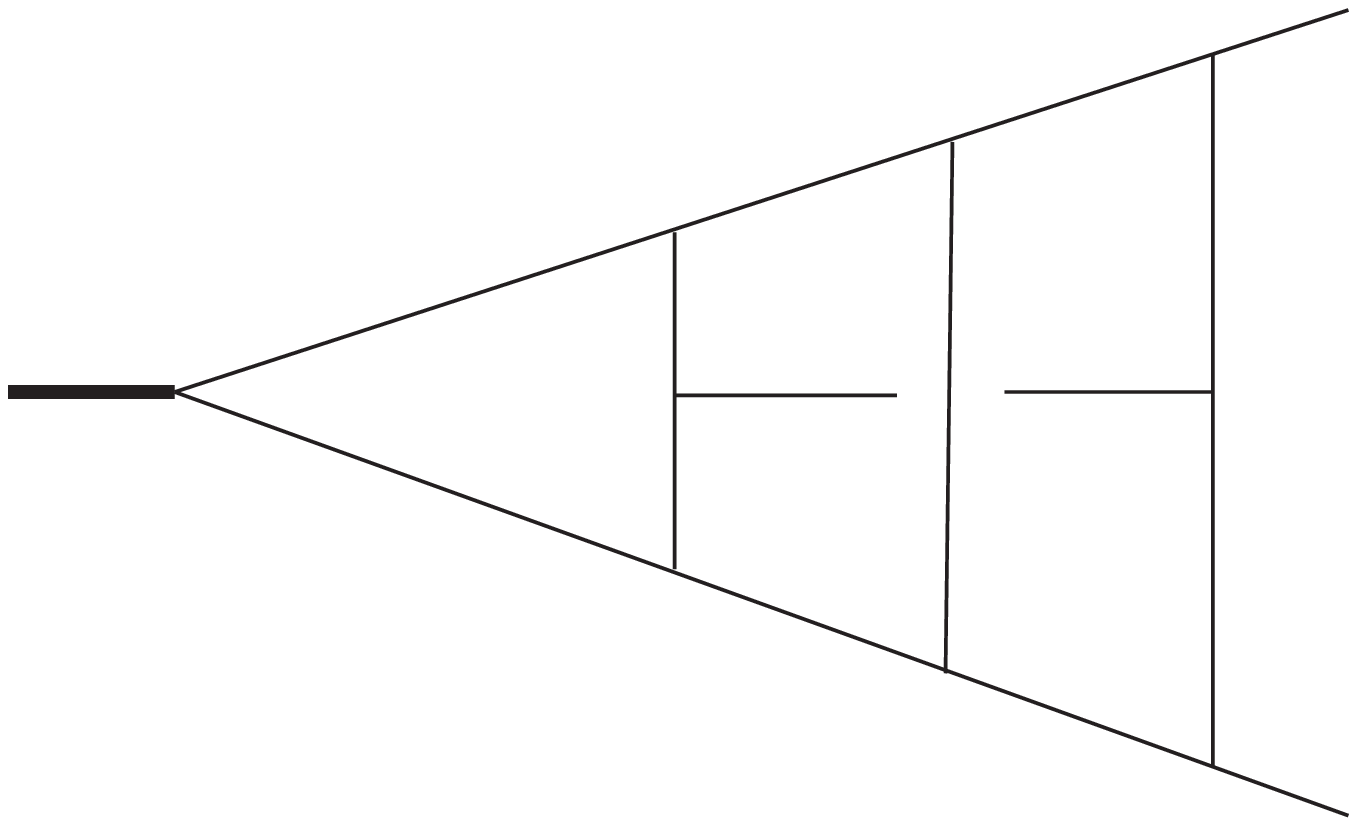} \qquad \Graph{.175}{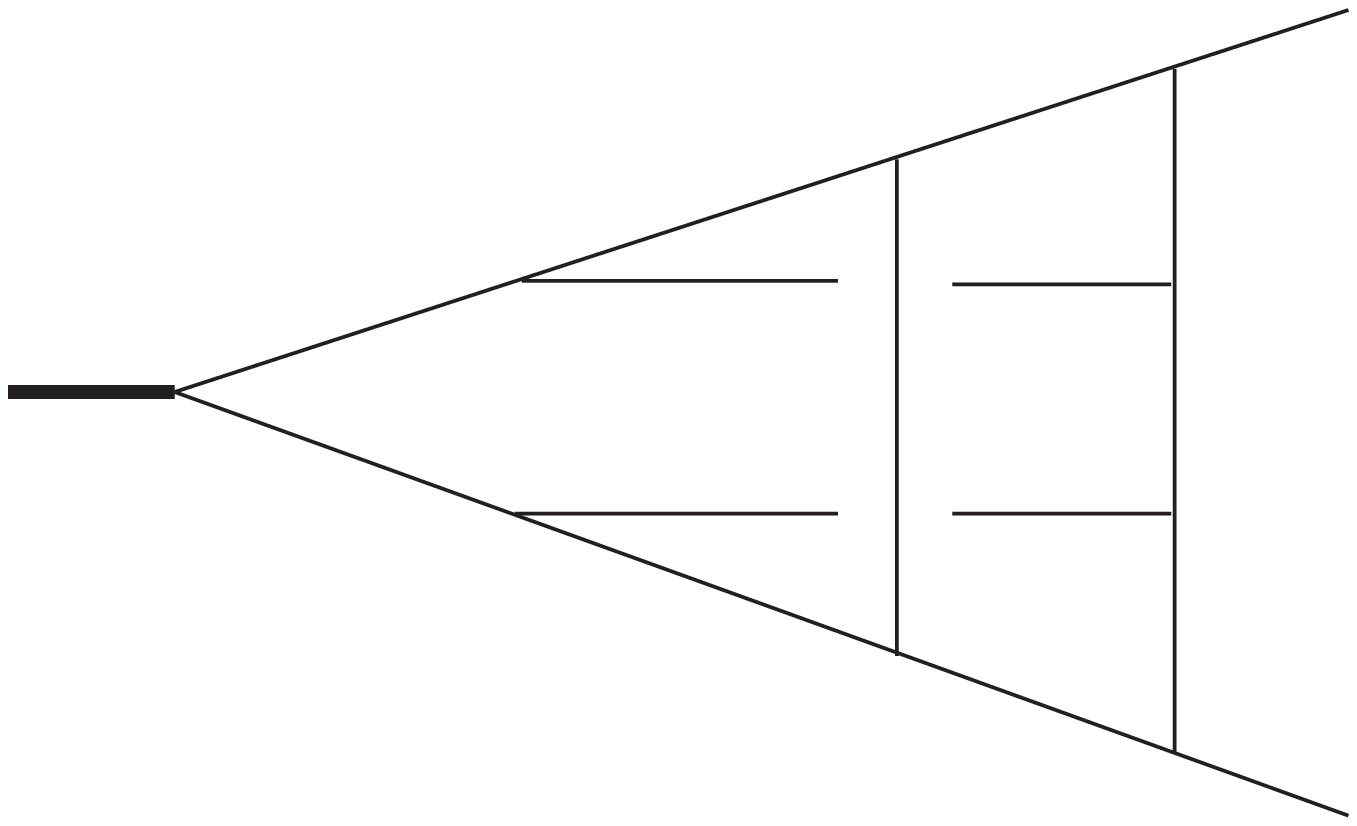} \qquad \Graph{.175}{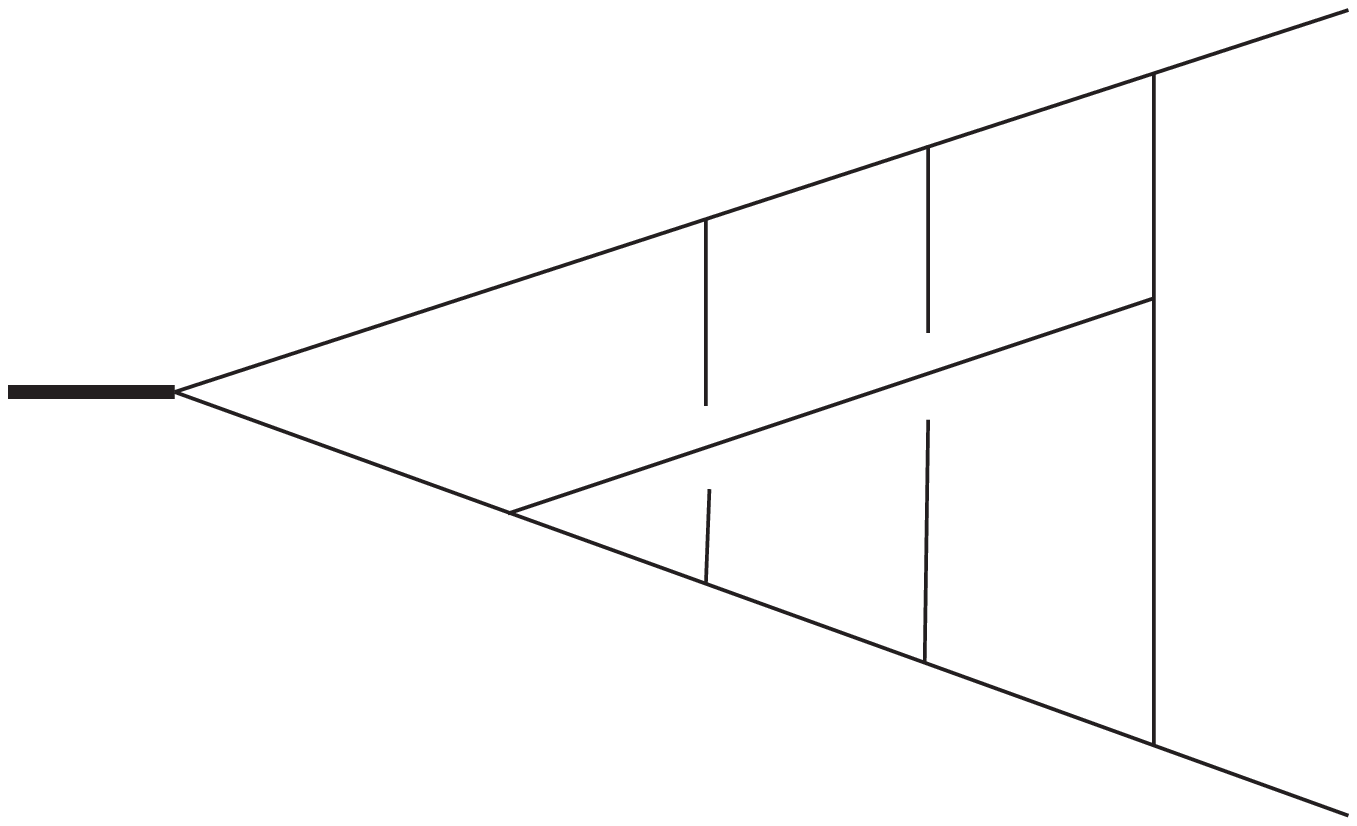} \qquad \Graph{.175}{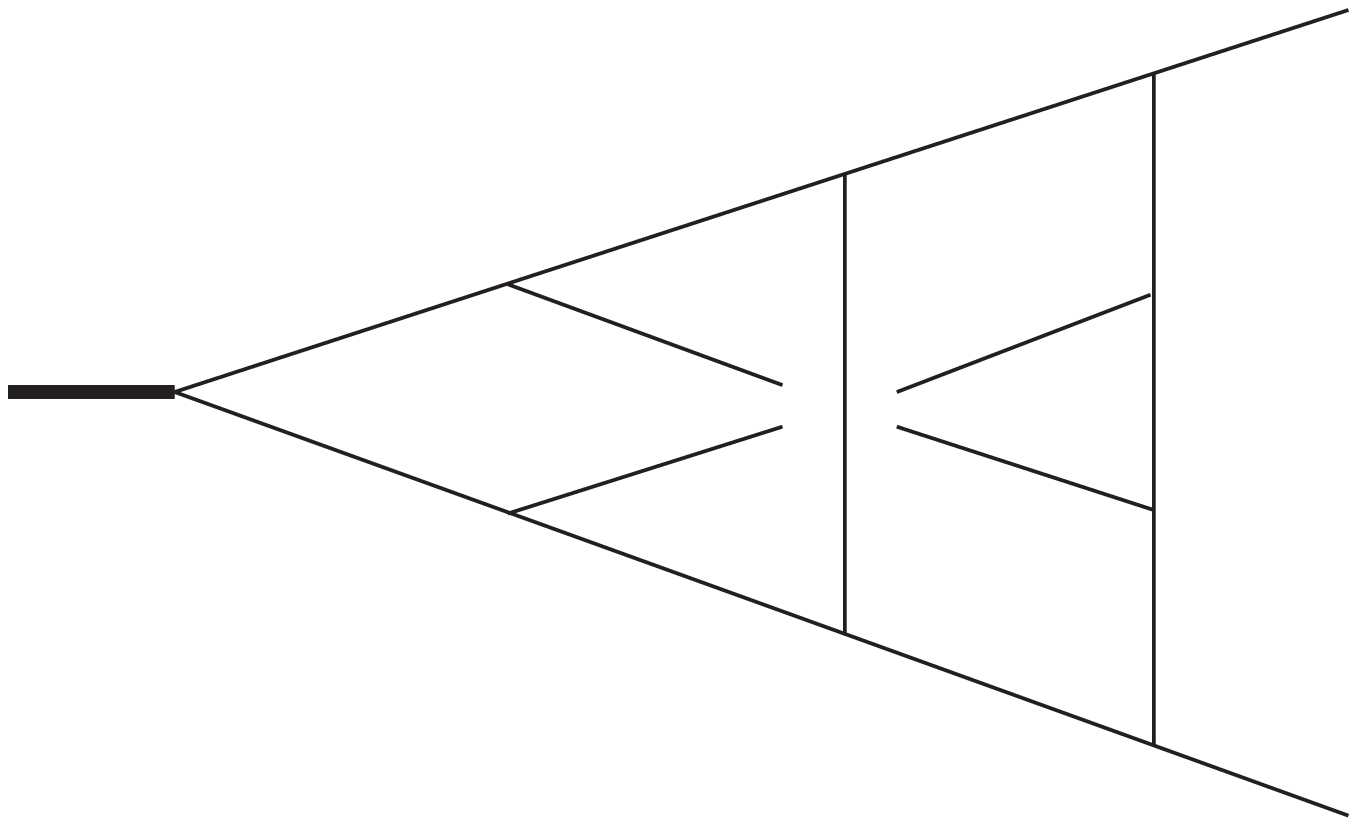}
\end{align*}
\caption{All irreducible top-level integral topologies occurring in our form factors.
Out of these, only the two framed topologies  eventually contribute to the $\epsilon^{-2}$ poles of the form factors and thus to the cusp anomalous dimensions for our choice of basis integrals.
The topologies in the first two rows are integrable directly in Feynman parameters with {\HyperInt}.
For all but the last two topologies, we found simple changes of variables that render them accessible (`linearly reducible') as well.%
}%
\label{fig:sampletopos}%
\end{figure*}

While the beta function of Quantum Chromodynamics (QCD) determines the running of the coupling due to ultraviolet divergences, the cusp anomalous dimensions of the quark and gluon determine the leading infrared singularities of massless scattering amplitudes \cite{Korchemsky:1985xj}.
To second order in the strong coupling constant, these anomalous dimensions were already known implicitly before the appearance of \cite{Korchemsky:1985xj} from investigations of the next-to-leading Dokshitzer-Gribov-Lipatov-Altarelli-Parisi (DGLAP) splitting functions \cite{Curci:1980uw,Furmanski:1980cm,Kalinowski:1980ju,Kalinowski:1980wea,Gunion:1984xw,Gunion:1985pg}.

The phenomenological relevance of the cusp anomalous dimensions to the resummation of prominent QCD observables is well-established, considered in some cases at the next-to-next-to-next-to-leading (four-loop) logarithm level a decade ago \cite{Becher:2008cf,Abbate:2010xh},
but their calculation to higher orders in QCD perturbation theory is a challenging task. After the completion of the two-loop calculations, roughly twenty years elapsed before the appearance of a first analytic calculation of the three-loop cusp anomalous dimensions \cite{Moch:2004pa,Vogt:2004mw} from the three-loop DGLAP splitting functions. Over the last few years, a number of approximate numerical \cite{Moch:2017uml,Moch:2018wjh} and partial analytic \cite{Grozin:2015kna,Henn:2016men,Ruijl:2016pkm,vonManteuffel:2016xki,Lee:2016ixa,Lee:2017mip,Grozin:2018vdn,Lee:2019zop,Henn:2019rmi,Bruser:2019auj,vonManteuffel:2019wbj} results have appeared at the four-loop level; just as for the beta function of massless QCD, now known to five-loop order after years of intensive investigation  \cite{Baikov:2016tgj,Herzog:2017ohr,Luthe:2017ttg,Chetyrkin:2017bjc}, a high degree of automation and significant computer resources enabled this progress.

Up to three loops, the cusp anomalous dimensions of the quark and gluon are related to each other by the quadratic Casimir scaling principle \cite{Korchemsky:1988si,Becher:2009cu,Gardi:2009qi,Becher:2009qa,Dixon:2009gx}. The authors of \cite{Moch:2018wjh} conjectured that this no longer holds at the four-loop level, but is rather {\it generalized} to accommodate novel color structures built out of quartic Casimir operators.
This generalized Casimir scaling proposal was recently corroborated by two independent theoretical studies \cite{Catani:2019rvy,Becher:2019avh}. Using further conjectural input from \cite{Bruser:2019auj}, an analytic form of the four-loop QCD cusp anomalous dimensions was put forward very recently in \cite{Henn:2019swt}. 

The primary goal of this Letter is to definitively calculate the four-loop QCD cusp anomalous dimensions analytically from first principles. We follow \cite{vonManteuffel:2015gxa} and extract the cusp anomalous dimensions from expansions of massless four-loop quark and gluon QCD form factors through to $\mathcal{O}\big(\epsilon^{-2}\big)$.
We employed new methods for the reduction to master integrals \cite{vonManteuffel:2014ixa,vonManteuffel:2016xki,vonManteuffel:2019wbj}, and we tailored the choice of master integrals to simplify their evaluation \cite{vonManteuffel:2014qoa,Schabinger:2018dyi,vonManteuffel:2019gpr}.

Our approach relies on the existence of a basis of integrals which are finite as $\epsilon \to 0$ and sufficiently well-behaved with respect to the transcendental weight filtration. 
In such a basis, many of the most complicated integral topologies (see Fig.~\ref{fig:sampletopos}) do not contribute to the higher-order poles in $\epsilon$. To illustrate this point, consider the result
\begin{align}
\label{eq:sampleint}
&\Graph{.185}{D_12_27631} =
\frac{7}{18\epsilon^8}
+\frac{55}{24\epsilon^7}
-\frac{1}{\epsilon^6}\bigg(
    \frac{67}{9} \zeta_2
    +\frac{797}{144}
\bigg)
\nonumber\\&
-\frac{1}{\epsilon^5}\bigg(
    \frac{442}{9} \zeta_3
    +\frac{643}{18} \zeta_2
    -\frac{1193}{144}
\bigg)
-\frac{1}{\epsilon^4}\bigg(
    \frac{9199}{360} \zeta_2^2
    +\frac{3547}{18}\zeta_3
\nonumber\\&
    -\frac{7793}{72}\zeta_2
    -\frac{1013}{48}
\bigg)
+\frac{1}{\epsilon^3}\bigg(
    \frac{27617}{36} \zeta_3 \zeta_2
    -\frac{2858}{3} \zeta_5
\nonumber\\&
    -\frac{3439}{180} \zeta_2^2
    +\frac{60893}{72}\zeta_3
    -\frac{1897}{8}\zeta_2
    -\frac{43895}{144}
\bigg)
\\&
+\frac{1}{\epsilon^2}\bigg(
    \frac{179927}{72} \zeta_3^2
    -\frac{40853}{252} \zeta_2^3
    -2780 \zeta_5
    +\frac{23467}{9} \zeta_3 \zeta_2
\nonumber\\&
    +\frac{132359}{180} \zeta_2^2
    -\frac{66607}{24} \zeta_3
    -\frac{5423}{72} \zeta_2
    +\frac{311383}{144}
\bigg)
+\mathcal{O}\big(\epsilon^{-1}\big),\nonumber
\end{align}
in $4-2\epsilon$ dimensions, given in the conventions of \cite{vonManteuffel:2019gpr}.
In our basis of finite integrals, this integral topology first contributes to the Laurent expansion at order $\epsilon^{-1}$. Hence, we were able to produce Eq.~\eqref{eq:sampleint} from the subtopologies without having to actually integrate any integral of the topology itself.

We also observe that for four-loop Feynman diagrams containing at least one closed fermion loop, our finite basis integrals in the two most complicated topologies (see Fig.~\ref{fig:sampletopos}) only contribute to the finite parts, $\bigO{\epsilon^0}$.
The remaining finite basis integrals can be integrated with the program {\HyperInt} \cite{Panzer:2014caa}, and so we obtained the complete matter dependence of the $\epsilon^{-1}$ poles of the form factors. It only involves zeta values of weight at most 6.

As a consequence, we are able to offer new results on another well-studied pair of quantities, the quark and gluon collinear anomalous dimensions  (see \cite{Becher:2014oda} for the QCD results up to three loops). Historically, the color dipole conjecture of \cite{Becher:2009cu,Gardi:2009qi,Becher:2009qa,Dixon:2009gx} offered an enticingly-simple prediction for the $\epsilon^{-1}$ poles of quite general massless QCD scattering amplitudes. In the color dipole picture, simple singular dressing factors for each external parton (quark or gluon jet functions) contribute poles of collinear origin to the $\epsilon^{-1}$ pole of the logarithm of the amplitude under consideration. The collinear anomalous dimensions fix the $\epsilon^{-1}$ poles of these jet functions.

While we now know that the $\epsilon^{-1}$ poles of QCD amplitudes receive color quadrupole corrections at three-loop order and beyond from the soft sector \cite{Caron-Huot:2013fea,Almelid:2015jia,Henn:2016jdu}, the four-loop collinear anomalous dimensions have long been of interest in planar $\mathcal{N} = 4$ super Yang-Mills theory \cite{Cachazo:2007ad,Dixon:2017nat}, where the dipole conjecture does appear to hold \cite{Alday:2007hr}. Four-loop collinear anomalous dimensions in QCD will not be needed for phenomenological purposes any time soon, but partial results for the quark case have nevertheless already appeared \cite{Henn:2016men,Lee:2016ixa,Lee:2017mip,Lee:2019zop}.
We give complete analytic results for the matter dependence of the four-loop quark {\it and} gluon collinear anomalous dimensions.

\section{Setup and Integral Reduction}

We study quantum corrections in massless QCD to decays of both photons {\it and} Higgs bosons, \textit{i.e.} the processes $\gamma^\ast(q)\rightarrow q(p_1)\bar{q}(p_2)$ and $h(q)\rightarrow g(p_1)g(p_2)$, with $p_1^2=p_2^2=0$  and $q^2=(p_1+p_2)^2$.
We define {\it form factors} by interfering the bare $L-$loop scattering amplitudes with the tree amplitudes, summing over polarizations and colors, and then normalizing to the corresponding tree-level results,
\begin{multline}\label{eq:expbareg}
\ff_{\rm bare}^r\left(\alpha_s^{\rm bare}, q^2, \mu_\epsilon^2, \epsilon\right) = \\
1 + \sum_{L = 1}^\infty
        \left(\frac{\alpha_s^{\rm bare}}{4\pi}\right)^L 
        \left(\frac{4\pi \mu_\epsilon^2}{-q^2 e^{\gamma_E}}\right)^{L \epsilon}
        \ff_L^r(\epsilon).
\end{multline}
Here and in what follows, $r = q$ or $g$. We work in conventional dimensional regularization with $\epsilon = (4-d)/2$ and expand in the bare
strong coupling constant, $\alpha_s^{\rm bare}$.
Further, $\mu_\epsilon$ denotes the 't~Hooft scale and $\gamma_E$ is Euler's constant.
We consider the color structures of the four-loop corrections and find for the bare quark form factor
\begin{widetext}
\begin{align}
\label{eq:cfdecompq}
\ff_4^{q}(\epsilon) = 
      \qFFterm{N_f^3 C_F}{F}
&   + \qFFterm{N_f^2 C_A C_F}{AF}
    + \qFFterm{N_f^2 C_F^2}{F2}
    + \qFFterm{N_{q \gamma} N_f \frac{d_F^{abc}d_F^{abc}}{N_F}}{d3FF}
    + \qFFterm{N_f \frac{d_F^{abcd}d_F^{abcd}}{N_F}}{d4FF}
\nonumber \\
    +\qFFterm{N_f C_A^2 C_F}{A2F}
&   +\qFFterm{N_f C_A C_F^2}{AF2}
    +\qFFterm{N_f C_F^3}{F3}
    +\qFFterm{N_{q \gamma} C_A \frac{d_F^{abc}d_F^{abc}}{N_F}}{Ad3FF}
    +\qFFterm{N_{q \gamma} C_F \frac{d_F^{abc}d_F^{abc}}{N_F}}{Fd3FF}
\nonumber \\
    +\qFFterm{C_A^3 C_F}{A3F}
&   +\qFFterm{C_A^2 C_F^2}{A2F2}
    +\qFFterm{C_A C_F^3}{AF3}
    +\qFFterm{C_F^4}{F4}
    +\qFFterm{\frac{d_A^{abcd}d_F^{abcd}}{N_F}}{d4AF}
    ,
\end{align}
and the bare gluon form factor,
\begin{align}
\label{eq:cfdecompg}
\ff_4^{g}(\epsilon) =
      \gFFterm{N_f^3 C_A}{A}
&   + \gFFterm{N_f^3 C_F}{F}
    + \gFFterm{N_f^2 C_A^2}{A2}
    + \gFFterm{N_f^2 C_A C_F}{AF}
    + \gFFterm{N_f^2 C_F^2}{F2}
    + \gFFterm{N_f^2 \frac{d_F^{abcd}d_F^{abcd}}{N_A}}{d4FF}
\nonumber \\
    + \gFFterm{N_f C_A^3}{A3}
&   + \gFFterm{N_f C_A^2 C_F}{A2F}
    + \gFFterm{N_f C_A C_F^2}{AF2}
    + \gFFterm{N_f C_F^3}{F3}
    + \gFFterm{N_f \frac{d_A^{abcd}d_F^{abcd}}{N_A}}{d4AF}
\nonumber \\
    + \gFFterm{C_A^4}{A4}
&   + \gFFterm{\frac{d_A^{abcd}d_A^{abcd}}{N_A}}{d4AA}.
\end{align}
\end{widetext}
We denote the number of light quark flavors by $N_f$ and their charge-weighted sum, normalized to the charge of the external quark $q$, by $N_{q\gamma}\equiv  {\sum_{q^\prime} e_{q^\prime}}/e_q$. The color decompositions \eqref{eq:cfdecompq} and \eqref{eq:cfdecompg} follow the notation and conventions of \cite{vanRitbergen:1998pn}; the {\tt Form} program {\tt Color.h} was used to carry out the color algebra assuming a theory with a general simple compact gauge group. For the case of $SU(N_c)$, we have $d_F^{abc}d_F^{abc}=(N_c^2-1)(N_c^2-4)/(16 N_c)$ and all other invariants are given in Eqs.~(12) and (13) of \cite{Czakon:2004bu}.
Without loss of generality, we set $T_F = 1/2$ throughout.

The color coefficients in Eqs.\ \eqref{eq:cfdecompq} and \eqref{eq:cfdecompg} are derived as follows. We generate the four-loop Feynman diagrams with the program {\tt QGraf} \cite{Nogueira:1991ex} and
obtain a total of 5728 (43220) diagrams for the quark (gluon) form factor.
We encounter 100 twelve-line top-level topologies and match all of the diagrams to just ten complete sets of eighteen denominators
 with {\tt Reduze\;2} \cite{Bauer:2000cp,Studerus:2009ye,vonManteuffel:2012np}.
Here, one such set of denominators may cover several twelve-line top-level topologies.
We carry out all numerator algebra with {\tt Form\;4} \cite{Kuipers:2012rf} and arrive at linear combinations of scalar Feynman integrals with up to six inverse propagators.

We exploit linear relations to reduce the integrals to master integrals, using the program {\Finred} by the first author.
For the reduction of the amplitude, we employ conventional momentum space integration by parts, Lorentz, and sector symmetry identities~\cite{Tkachov:1981wb,Chetyrkin:1981qh,Gehrmann:1999as,Laporta:2001dd}. In some instances, we also found it useful to apply syzygy technology \cite{Gluza:2010ws,Schabinger:2011dz,Ita:2015tya,Larsen:2015ped,Boehm:2017wjc}.
For the basis change to finite integrals, we made heavy use of first- and second-order annihilators \cite{Lee:2013mka,Bitoun:2017nre} in the Lee-Pomeransky representation~\cite{Lee:2013hzt}.
Instead of resorting to computer algebra systems, we
compute syzygies with linear algebra \cite{CabarcasDing,Schabinger:2011dz} using {\Finred} as a linear solver.
This method allows us to reduce integrals with high powers of propagators and no numerators.

The program {\Finred} implements finite field sampling and rational reconstruction \cite{vonManteuffel:2014ixa,vonManteuffel:2016xki}
and supports distributed computations to efficiently solve large linear systems.
We solved sectors with more than $10^8$ equations and reconstructed identities from up to $\bigO{40}$ 64-bit-based prime fields and $\bigO{600}$ values for the space-time dimension.
Including the identities for basis changes and dimensional shifts, our compressed reduction tables consume $\bigO{10\,\text{TB}}$ on disk.

In total, we find 294 master integrals.
Twenty of the top-level topologies turn out to be irreducible, see Fig.~\ref{fig:sampletopos}, with up to four master integrals per topology.
We would like to point out an interesting relation between master integrals of three distinct nine-line topologies,
\begin{align}
\label{eq:cross-sector-ibp}
\Graph{.19}{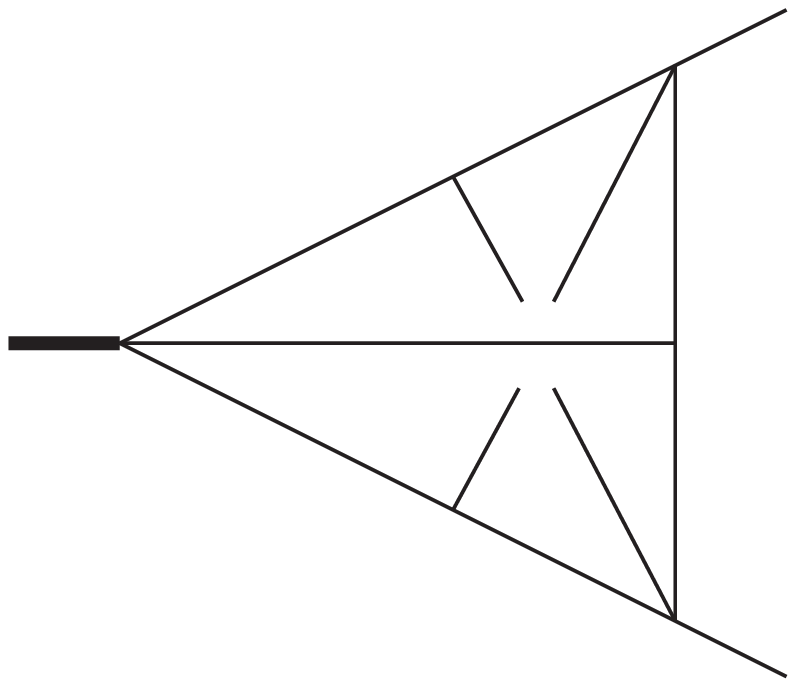} &=
\frac{4(2d-7)}{3d-11}
\Graph{.19}{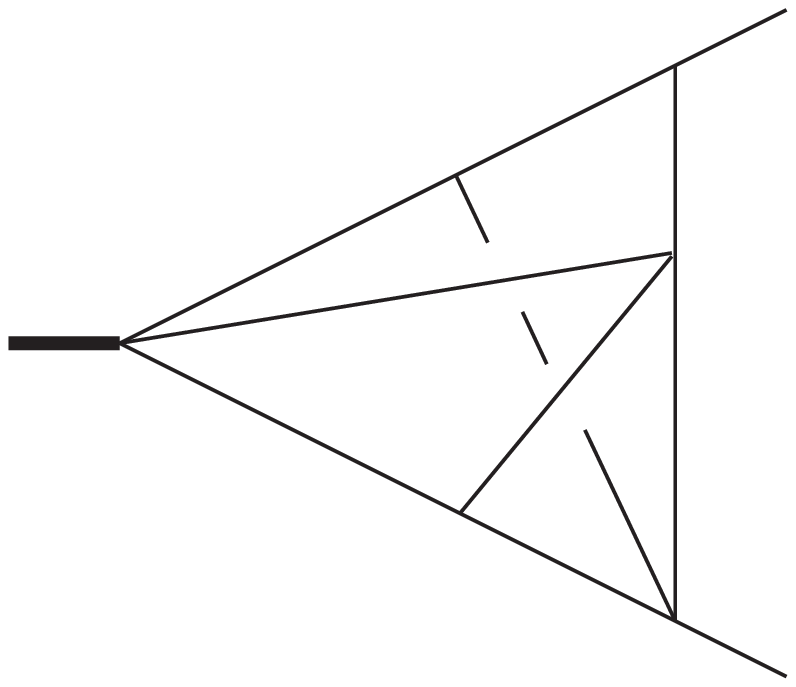}
+ \frac{5(5-d)}{3 d -11}
\Graph{.19}{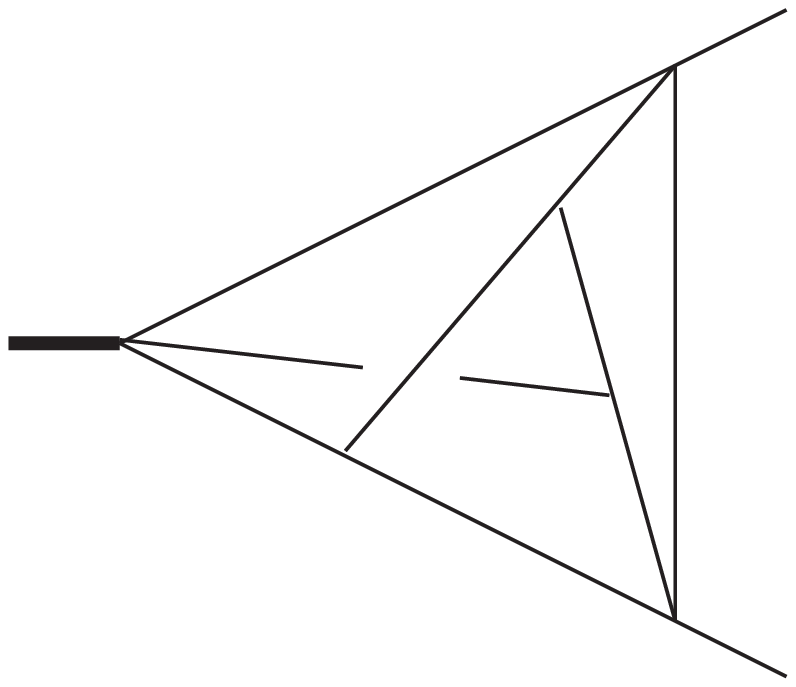}\notag\\
&\quad + \text{subsectors},
\end{align}
which can be obtained from a common parent topology.

As a check of our reductions, we calculated the matter dependent contributions to the quark and gluon form factors in a general $R_\xi$ gauge and verified explicitly that terms proportional to $\xi$ cancel. This cancellation occurred only after accounting for relations between color invariants (for general Lie algebras) and Eq.\ \eqref{eq:cross-sector-ibp}.

\section{Results}

We insert our analytical solutions for the finite master integrals to obtain $\epsilon$-expanded expressions for the form factors.
Our main new results are extracted from the $\mathcal{O}\big(\epsilon^{-1}\big)$ coefficients proportional to $N_f$ or $N_{q \gamma}$ on the second lines of Eqs.\ \eqref{eq:cfdecompq} and \eqref{eq:cfdecompg}, and the $\mathcal{O}\big(\epsilon^{-2}\big)$ coefficients of the color factors on the second and third lines of Eqs.\ \eqref{eq:cfdecompq} and \eqref{eq:cfdecompg}. Explicit expressions for these coefficients are provided in the appendix. The coefficients on the first lines of Eqs.\ \eqref{eq:cfdecompq} and \eqref{eq:cfdecompg} have already been calculated through to $\mathcal{O}\big(\epsilon^0\big)$ by a subset of the authors \cite{vonManteuffel:2016xki,vonManteuffel:2019wbj} and others \cite{Henn:2016men,Lee:2017mip,Lee:2019zop}. For the sake of completeness, we also calculated the quartic color coefficient $\qFF{d4FF}$ through to $\mathcal{O}\big(\epsilon^0\big)$ ourselves to confirm the result of \cite{Lee:2019zop}. 
 
 We determine the four-loop cusp anomalous dimensions in the framework of \cite{Moch:2005id} from the $\epsilon^{-2}$ poles. Our results for $\Gamma^q_4$ and $\Gamma^g_4$ confirm the generalized Casimir scaling principle, so that we can write them together as
\begin{widetext}
\begin{align}
\label{eq:QCDcusp}
\Gamma^r_4 &= 
\CS{N_f^3 C_R} \left(\frac{64}{27} \zeta _3-\frac{32}{81}\right)
+ \CS{N_f^2 C_A C_R} \left(-\frac{224}{15} \zeta _2^2+\frac{2240}{27} \zeta _3-\frac{608}{81} \zeta _2+\frac{923}{81}\right) 
+ \CS{N_f^2 C_F C_R} \left(\frac{64}{5} \zeta _2^2-\frac{640}{9} \zeta _3+\frac{2392}{81}\right)
\nonumber \\ &
+ \CS{N_f C_A^2 C_R} \left(
    \frac{2096}{9} \zeta _5
    +\frac{448}{3} \zeta _3 \zeta _2
    -\frac{352}{15} \zeta _2^2
    -\frac{23104}{27} \zeta _3
    +\frac{20320}{81} \zeta _2
    -\frac{24137}{81}
\right)
\nonumber\\&
+\CS{N_f C_A C_F C_R} \left(
    160 \zeta _5
    -128 \zeta _3 \zeta _2
    -\frac{352}{5} \zeta _2^2 
    +\frac{3712}{9} \zeta _3
    +\frac{440}{3} \zeta _2
    -\frac{34066}{81}
\right) 
+ \CS{N_f C_F^2 C_R}  \left(
    -320 \zeta _5
    +\frac{592}{3} \zeta _3
    +\frac{572}{9}
\right)
\nonumber\\&
+ \CS{N_f \frac{d^{abcd}_F d^{abcd}_R}{N_R}} \left(
    -\frac{1280}{3} \zeta _5
    -\frac{256}{3} \zeta _3
    +256 \zeta _2
\right)
+ \CS{\frac{d^{abcd}_A d^{abcd}_R}{N_R}} \left(
    -384 \zeta _3^2
    -\frac{7936}{35} \zeta _2^3
    +\frac{3520}{3} \zeta _5
    +\frac{128}{3} \zeta _3
    -128 \zeta _2
\right)
\nonumber\\&
+\CS{C_A^3 C_R} \left(
    -16 \zeta _3^2
    -\frac{20032}{105} \zeta _2^3
    -\frac{3608}{9} \zeta _5
    -\frac{352}{3} \zeta _3 \zeta _2
    +\frac{3608}{5} \zeta _2^2
    +\frac{20944}{27} \zeta _3
    -\frac{88400}{81} \zeta _2
    +\frac{84278}{81}
\right),
\end{align}
\end{widetext}
where $R = F$ if $r = q$ and $R = A$ if $r = g$.
Only color structures expected from the Wilson loop picture appear.

Our result in Eq.~\eqref{eq:QCDcusp} agrees with Eq.~(6.3) of \cite{Henn:2019swt}, which is based on a calculation of the cusp anomalous dimension of $\mathcal{N} = 4$ super Yang-Mills theory, a supersymmetric decomposition \cite{Grozin:2015kna,Henn:2019rmi}, generalized Casimir scaling, see also \cite{Moch:2018wjh,Catani:2019rvy,Becher:2019avh},
a conjecture for the $N_f C_A C_F C_R$ term \cite{Bruser:2019auj},
and previously known results for the other $N_f$-dependent terms \cite{Grozin:2015kna,Ruijl:2016pkm,vonManteuffel:2016xki,Lee:2017mip,Grozin:2018vdn,Lee:2019zop,Henn:2019rmi,vonManteuffel:2019wbj}.

Another strong independent check is given by comparing our result to the approximate numerical analysis of \cite{Moch:2017uml,Moch:2018wjh}. We find very solid agreement with Table 1 of \cite{Moch:2018wjh}, suggesting that their error estimates were actually conservative.

It was pointed out already in \cite{Henn:2019swt} that Eq.~\eqref{eq:QCDcusp} correctly predicts the four-loop cusp anomalous dimension of the $\mathcal{N} = 4$ model \cite{Henn:2019swt,Huber:2019fxe} through the principle of maximal transcendentality \cite{Kotikov:2001sc,Kotikov:2002ab}. Indeed, in the notation of this Letter, Eq.~(6.1) of \cite{Henn:2019swt} takes the form
\begin{align*}
\Gamma^{\mathcal{N} = 4}_4 
&= \CS{\frac{d^{abcd}_A d^{abcd}_A}{N_A}} \left(-384 \zeta _3^2-\frac{7936}{35} \zeta_2^3\right) \\
&+ \CS{C_A^4} \left(-16 \zeta _3^2 - \frac{20032}{105} \zeta _2^3\right),
\end{align*}
which matches precisely the terms of Eq.~\eqref{eq:QCDcusp} that have the highest transcendental weight (6).

The collinear anomalous dimensions can be read off from the $\epsilon^{-1}$ poles of the logarithm of the renormalized form factors. We confirmed that all higher order poles are as predicted by Eq.~(6.22) from \cite{Gehrmann:2010ue} in terms of the cusp and lower-loop collinear anomalous dimensions, together with the coefficients $\beta_{L-1}$ of the beta function \cite{vanRitbergen:1997va,Czakon:2004bu}.
Equivalently, the collinear anomalous dimensions are defined as
\begin{align}
\label{eq:colqgdef}
\gamma_4^r &= G^r_4[0] - \beta_0 G^r_3[1] - \beta_1 G^r_2[1] - \beta_2 G^r_1[1]
\nonumber \\&
 \vphantom{\Big(}+ \beta_0^2 G^r_2[2] + 2 \beta_0 \beta_1 G^r_1[2] - \beta_0^3 G^r_1[3] + 8 \beta_3 \delta_{g r}
\end{align}
in the framework of \cite{Moch:2005id} (see also Eq.~(20) of \cite{Ravindran:2006cg}). Here, $G^r_L[k]$ denotes the coefficient of $\epsilon^k$ in the series $G^r_L(\epsilon)$ defined in Eqs.~(2.14)--(2.17) of \cite{Moch:2005id} in terms of the bare form factors. These coefficients can be extracted from the four-loop expansions in our appendix, together with the well-known higher orders in $\epsilon$ of the bare one-, two-, and three-loop form factors given in \cite{Lee:2010cga,Gehrmann:2010ue,Gehrmann:2005pd,Gehrmann:2010tu} or \cite{vonManteuffel:2015gxa}.

Note that our $\gamma_4^q$ and $\gamma_4^g$ are $(-2)$ times the four-loop collinear anomalous dimensions defined in \cite{Becher:2014oda}, and \cite{Henn:2016men,Lee:2016ixa,Lee:2017mip,Lee:2019zop} follow \cite{Becher:2014oda} rather than \cite{Moch:2005id,Falcioni:2019nxk,Moch:2005tm}. We find
\begin{widetext}
\begin{align}
\label{eq:QCDcolq}
\gamma^q_4 &= 
\CS{N_f^3 C_F} \bigg(\frac{128}{135}\zeta_2^2+\frac{1424}{243} \zeta _3+\frac{16}{27} \zeta_2-\frac{37382}{6561}\bigg)
+ \CS{N_f^2 C_F^2} \left(
    \frac{1040}{9} \zeta_5
    -\frac{224}{9} \zeta_3 \zeta_2
    -\frac{8032}{135} \zeta_2^2
    -\frac{4232}{81} \zeta_3
    +\frac{1972}{27} \zeta_2
    +\frac{9965}{486}
\right) 
\nonumber\\&
+ \CS{N_f^2 C_A C_F} \left(
    -\frac{1184}{9} \zeta _5
    +\frac{256}{9} \zeta _3 \zeta _2
    +\frac{152}{15} \zeta _2^2
    +\frac{14872}{243} \zeta _3
    +\frac{41579}{729} \zeta_2
    -\frac{97189}{17496}
\right) 
\nonumber\\&
+ \CS{N_f C_A^2 C_F}\left(
    \frac{6916}{9} \zeta _3^2
    +\frac{24184}{315} \zeta _2^3
    +\frac{6088}{27} \zeta_5
    -\frac{3584}{9} \zeta_3 \zeta_2
    -\frac{17164}{45} \zeta _2^2
    +\frac{140632}{243} \zeta_3
    -\frac{445117}{729} \zeta_2
    +\frac{326863}{1944}
\right)
\nonumber\\&
+  \CS{N_f C_A C_F^2} \left(
    -\frac{3400}{3} \zeta _3^2
    +\frac{5744}{35} \zeta _2^3
    -\frac{4472}{3} \zeta _5
    +\frac{3904}{9} \zeta _3 \zeta_2
    +\frac{105488}{135} \zeta _2^2
    -\frac{23518}{81} \zeta_3
    +\frac{673}{27} \zeta _2
    -\frac{1092511}{972}
\right)
\nonumber\\&
+ \CS{N_f C_F^3}  \left(
    368 \zeta _3^2
    -\frac{117344}{315} \zeta _2^3
    +\frac{3872}{3} \zeta _5
    -\frac{512}{3} \zeta _3 \zeta _2
    -\frac{668}{5} \zeta _2^2
    -\frac{1120}{9} \zeta _3
    +322\zeta _2
    +\frac{27949}{108}
\right) 
\nonumber\\&
+ \CS{N_f \frac{d^{abcd}_F d^{abcd}_F}{N_F}} \left(
    \frac{1216}{3} \zeta _3^2
    +\frac{9472}{315} \zeta _2^3
    -\frac{21760}{9} \zeta_5
    +128 \zeta _3 \zeta _2
    -\frac{320}{3} \zeta _2^2
    -\frac{5312}{9} \zeta _3
    +\frac{4544}{3} \zeta_2
    -384
\right)
+\bigO{N_f^0}
\end{align}
for the matter-dependent parts of the quark collinear anomalous dimension and
\begin{align}
\label{eq:QCDcolg}
&\gamma^g_4 = \CS{N_f^3 C_A} \left(\frac{256}{135}\zeta_2^2-\frac{400}{243} \zeta _3-\frac{16}{81} \zeta_2-\frac{15890}{6561}\right)+\CS{N_f^3 C_F} \left(\frac{308}{243}\right)
+ \CS{N_f^2 \frac{d^{abcd}_F d^{abcd}_F}{N_A}}\left(\frac{1024}{3} \zeta _3 -\frac{1408}{9}\right)
\nonumber\\&
+ \CS{N_f^2 C_A^2} \left(-\frac{1024}{9} \zeta _5 - 32 \zeta _3 \zeta _2+\frac{3128}{135} \zeta _2^2+\frac{37354}{243} \zeta _3-\frac{13483}{729} \zeta_2 + \frac{611939}{17496}\right)+\CS{N_f^2 C_A C_F} \left(\frac{304}{9} \zeta_5+\frac{32}{3} \zeta _3 \zeta _2+\frac{128}{45} \zeta _2^2
\right.\nonumber\\&\left.
-\frac{1688}{81} \zeta_3-\frac{172}{9} \zeta_2+\frac{1199}{18}\right)
+\CS{N_f^2 C_F^2} \left(-\frac{352}{9} \zeta_3+\frac{676}{27} \right) + \CS{N_f C_A^3}\left(-\frac{596}{9} \zeta _3^2+\frac{148976}{945} \zeta _2^3+\frac{16066}{27} \zeta_5+148 \zeta_3 \zeta_2
\right.\nonumber\\&\left.
-\frac{69502}{135} \zeta _2^2-\frac{260822}{243} \zeta_3+\frac{155273}{729} \zeta_2-\frac{421325}{1944}\right)+  \CS{N_f C_A^2 C_F} \left(152 \zeta _3^2+\frac{5632}{315} \zeta _2^3+\frac{8}{9} \zeta _5-176 \zeta _3 \zeta_2-\frac{1196}{45} \zeta _2^2+\frac{29606}{81} \zeta_3
\right.\nonumber\\&\left.
+\frac{3023}{9} \zeta _2-\frac{903983}{972}\right) + \CS{N_f C_A C_F^2}  \left( -80\zeta _3^2-\frac{320}{7} \zeta _2^3-\frac{1600}{3} \zeta _5+\frac{148}{5} \zeta _2^2+\frac{1592}{3} \zeta _3-2
   \zeta _2+\frac{685}{12}\right)+ \CS{N_f C_F^3}  \Big(46\Big)
\nonumber\\&
+ \CS{N_f \frac{d^{abcd}_A d^{abcd}_F}{N_A}} \left(\frac{1216}{3} \zeta _3^2-\frac{14464}{315} \zeta _2^3
-\frac{30880}{9} \zeta_5+1216 \zeta _3 \zeta _2+\frac{2464}{15}\zeta_2^2+\frac{2560}{9} \zeta _3 -64 \zeta_2+\frac{448}{9}\right)
+\bigO{N_f^0}
\end{align}
\end{widetext}
for the gluon collinear anomalous dimension. Several useful cross-checks on Eq.\ \eqref{eq:QCDcolq} exist, including results for the $\mathcal{O}\big(N_f^3\big)$ and $\mathcal{O}\big(N_f^2\big)$ terms of $\gamma^q_4$ \cite{Henn:2016men,Lee:2017mip} (see also \cite{vonManteuffel:2016xki,vonManteuffel:2019wbj}), the result of \cite{Lee:2019zop} for the coefficient of the quartic Casimir invariant in Eq.\ \eqref{eq:QCDcolq}, the result of \cite{Lee:2016ixa} for the large-$N_c$ limit of the $\mathcal{O}\big(N_f\big)$ terms of $\gamma^q_4$, and the very recent numerical estimate of the $\epsilon^{-1}$ pole of the full four-loop quark form factor \cite{Das:2019btv}. Our results are in perfect agreement with the available literature. Furthermore, we note that the three color structures proportional to $N_{q \gamma}$ drop out of Eq.~\eqref{eq:QCDcolq} in a non-trivial way.

While no independent results for $\gamma^g_4$ are immediately available, \cite{Davies:2016jie} provides the $\mathcal{O}\big(N_f^3\big)$ part of the four-loop virtual anomalous dimensions, $B_4^r$, allowing for an alternative extraction of the first two terms of Eq.\ \eqref{eq:QCDcolg} from the relation \cite{Ravindran:2004mb,Dixon:2008gr}
\begin{equation}
\label{eq:colfromvirt}
\gamma_4^r = 2 B_4^r + f_4^r,
\end{equation}
where $f_4^r$ denotes the four-loop eikonal anomalous dimensions of massless QCD.
Relating $f_4^g$ to $f_4^q$ by Casimir scaling, we obtain the relevant terms in $\gamma_4^g$ from that in $\gamma_4^q$ and find agreement with the direct calculation.

We observe that {\it all} expansion coefficients of our finite basis integrals which contribute to the $\epsilon^{-1}$ poles of the matter-dependent color structures from the first two lines of Eqs.\ \eqref{eq:cfdecompq} and \eqref{eq:cfdecompg} also contribute to the $\epsilon^{-2}$ poles. The checks mentioned earlier for the cusp anomalous dimensions, therefore, also test our results for both collinear anomalous dimensions.

Our results for the form factors, anomalous dimensions, and the $G^r_L(\epsilon)$ functions are provided in {\tt Mathematica} and {\tt Maple} format \cite{SuppMat}.

\section{Summary}
\vspace*{-1ex}

We presented the first complete, {\it ab initio} analytic calculation of the four-loop quark and gluon cusp anomalous dimensions.
In contrast to previous analytic work on the subject, our extraction of the gluon cusp anomalous dimension did not rely on any conjectured property of the cusp anomalous dimensions. It therefore also provides a direct analytic confirmation of the generalized Casimir scaling principle \cite{Moch:2018wjh,Catani:2019rvy,Becher:2019avh} at the four-loop level.
Finally, we presented the full analytic matter dependence of the four-loop quark and gluon collinear anomalous dimensions.

\noindent {\em Acknowledgments:}

We gratefully acknowledge Rutger Boels, Tobias Huber, and Gang Yang for useful discussions and collaborations on closely-related topics.
A.\ v.\ M.\ and R.\ M.\ S.\ gratefully acknowledge  Vladimir A.\ Smirnov for an illuminating discussion regarding the appearance of anomalous integration by parts identities. 
We are indebted to Hubert Spiesberger for essential help and the Precision Physics, Fundamental Interactions and Structure of Matter (PRISMA$^+$) excellence cluster for generous financial support with computing resources.
Our computations were carried out in part on the supercomputer Mogon at Johannes Gutenberg University Mainz, 
and we wish to thank the Mogon team for their technical support.
We thank Dalibor Djukanovic for generously providing additional
computing resources at the Helmholtz Institute at Johannes Gutenberg University Mainz.
We also thank the mathematical institutes of the University of Oxford and Humboldt University of Berlin for computing resources.
Substantial computing resources were provided by the
High Performance Computing Center at Michigan State University,
and we gratefully acknowledge the High Precision Computing Center team for their help and support.
This work was supported in part by the National Science Foundation under Grant No.~1719863.
Our figures were generated using {\tt JaxoDraw} \cite{Binosi:2003yf}, based on {\tt AxoDraw} \cite{Vermaseren:1994je}. Finally, we thank J.\ M.\ Henn, G.\ P.\ Korchemsky, B.\ Mistlberger, S.-O.\ Moch, and V.\ Ravindran for useful feedback on our manuscript.

\vspace{-2ex}
\appendix*
\section{APPENDIX}
\vspace{-1ex}
In this appendix, we provide the explicit $\epsilon$ expansions of the previously-unknown color coefficients used to derive our main results, Eqs.\ \eqref{eq:QCDcusp}, \eqref{eq:QCDcolq}, and \eqref{eq:QCDcolg}. We have
\begin{align}
\label{eq:ffquarkA2F}
& \qFF{A2F} =
\frac{121}{36\epsilon^5}
+\frac{1}{\epsilon^4}\left(
    \frac{4613}{108}
    -\frac{11}{3} \zeta_2
\right)
+\frac{1}{\epsilon^3}\left(
    \frac{11}{5} \zeta_2^2
    -\frac{242}{3} \zeta_3
\right.\nonumber\\&\left.
    +\frac{173}{9} \zeta_2
    +\frac{454943}{1296}
\right)
+\frac{1}{\epsilon^2}\left(
    \frac{1093}{18} \zeta_5
    +10\zeta_3 \zeta_2
    -\frac{1313}{15} \zeta_2^2
\right.\nonumber\\&\left.
    -\frac{24721}{27} \zeta_3+\frac{24313}{54} \zeta _2
    +\frac{18167275}{7776}
\right)
+\frac{1}{\epsilon}\left(
    \frac{2815}{18} \zeta_3^2
\right.\nonumber\\&
    +\frac{7493}{135} \zeta_2^3
    -\frac{18749}{27} \zeta_5
    +\frac{2000}{9} \zeta_3 \zeta_2
    +\frac{2822933}{648} \zeta_2
\nonumber\\&\left.
    -\frac{638216}{81} \zeta_3
    -\frac{24118}{45} \zeta_2^2
    +\frac{642302539}{46656}
\right)
+\mathcal{O}\left(\epsilon^0\right)\!
\\&
\label{eq:ffquarkAF2}
\qFF{AF2} =
-\frac{451}{162\epsilon^6}
+\frac{1}{\epsilon^5}\left(
    \frac{41}{27} \zeta_2
    -\frac{6740}{243}
\right)
+\frac{1}{\epsilon^4}\left(
    \frac{629}{27} \zeta_3
\right.\nonumber\\&\left.
    -\frac{373}{81} \zeta_2
    -\frac{45430}{243}
\right)
+\frac{1}{\epsilon^3}\left(
    \frac{2737}{135} \zeta_2^2
    +\frac{75875}{243}\zeta_3
    -\frac{3725}{27}\zeta_2
\right.\nonumber\\&\left.
    -\frac{17699357}{17496}
\right)
+\frac{1}{\epsilon^2}\left(
    \frac{632}{3} \zeta_5
    -\frac{7328}{81} \zeta_3 \zeta_2
    +\frac{75967}{405} \zeta_2^2
\right.\nonumber\\&\left.
    +\frac{3542089}{1458} \zeta_3
    -\frac{944588}{729} \zeta_2
    -\frac{503448835}{104976}
\right)
+\frac{1}{\epsilon}\left(
    \frac{8954}{315}\zeta_2^3
\right.\nonumber\\&\left.
    -\frac{81152}{81}\zeta_3^2
    +\frac{860897}{405}\zeta_5
    -\frac{9668}{243} \zeta_3 \zeta_2
    +\frac{653551}{810} \zeta_2^2
\right.\nonumber\\&\left.
    +\frac{12023774}{729} \zeta_3
    -\frac{38449933}{4374} \zeta _2
    -\frac{1463037025}{69984}
\right)
+\mathcal{O}\left(\epsilon^0\right)
\\&
\label{eq:ffquarkF3}
\qFF{F3} = 
\frac{2}{3\epsilon^7}
+\frac{46}{9\epsilon^6}
+\frac{1555}{54\epsilon^5}
+\frac{1}{\epsilon^4}\bigg(
    \frac{41473}{324}
    +\frac{38}{3}\zeta_2
    -\frac{92}{3} \zeta_3
\bigg)
\nonumber\\&
+\frac{1}{\epsilon^3}\left(
    -\frac{844}{45} \zeta_2^2
    -\frac{2002}{9} \zeta_3
    +\frac{361}{3} \zeta_2
    +\frac{482393}{972}
\right)
\nonumber\\&
+\frac{1}{\epsilon^2}\left(
    \frac{736}{9} \zeta_3 \zeta_2
    -\frac{4234}{45} \zeta_5
    -\frac{10358}{135} \zeta_2^2
    -\frac{84533}{54} \zeta_3
\right.\nonumber\\&\left.
    +\frac{13903}{18} \zeta_2
    +\frac{20199679}{11664}
\right)
+\frac{1}{\epsilon}\left(
    \frac{36410}{27} \zeta _3^2
    +\frac{169268}{945} \zeta_2^3
\right.\nonumber\\&\left.
    -\frac{185222}{135} \zeta _5
    -\frac{232}{27} \zeta_3 \zeta_2
    -\frac{277277}{810} \zeta_2^2
    -\frac{722675}{81} \zeta_3
\right.\nonumber\\&\left.
    +\frac{111242}{27} \zeta _2
    +\frac{198160061}{34992}
\right)
+\mathcal{O}\left(\epsilon^0\right)
\\&
\qFF{A3F} = 
-\frac{1331}{216\epsilon^5}
+\frac{1}{\epsilon^4}\left(\frac{121}{12}\zeta_2-\frac{5863}{72}\right)
-\frac{1}{\epsilon^3}\left(
    \frac{121}{10} \zeta_2^2
\right.\nonumber\\&\left.
    -\frac{3025}{12} \zeta_3
    +\frac{1927}{108} \zeta_2
    +\frac{73901}{108}
\right)
+\frac{1}{\epsilon^2}\left(
    \frac{1}{2}\zeta_3^2
    +\frac{626}{105} \zeta_2^3
\right.\nonumber\\&\left.
    -\frac{13013}{36} \zeta_5
    -77 \zeta_3 \zeta_2
    +\frac{2893}{12} \zeta_2^2
    +\frac{223525}{81} \zeta_3
    -\frac{31177}{36} \zeta_2
\right.\nonumber\\&\left.
    -\frac{27082243}{5832}
\right)
+\mathcal{O}\left(\epsilon^{-1}\right)
\\&
\qFF{A2F2} = 
\frac{4961}{648\epsilon^6}
+\frac{1}{\epsilon^5}\left(
    \frac{38657}{486}
    -\frac{451}{54}\zeta_2
\right)
+\frac{1}{\epsilon^4}\left(
    \frac{397}{90}\zeta_2^2
\right.\nonumber\\&\left.
    -\frac{7975}{54}\zeta_3
    -\frac{179}{36}\zeta_2
    +\frac{1066517}{1944}
\right)
+\frac{1}{\epsilon^3}\bigg(
    \frac{272}{3} \zeta_5
    +\frac{293}{9} \zeta_3 \zeta_2
\nonumber\\&\left.
    -\frac{31259}{270} \zeta_2^2
    -\frac{754991}{486} \zeta_3
    +\frac{358559}{972} \zeta_2
    +\frac{107084117}{34992}
\right)
\nonumber\\&
+\frac{1}{\epsilon^2}\left(
    \frac{6065}{18} \zeta_3^2
    +\frac{67988}{945} \zeta_2^3
    -\frac{16903}{18} \zeta_5
    +\frac{24518}{81} \zeta_3\zeta_2
\right.\nonumber\\&
    -\frac{281117}{270} \zeta_2^2
    -\frac{32023111}{2916} \zeta_3
    +\frac{23659021}{5832} \zeta_2
\nonumber\\&\left.
    +\frac{1585980203}{104976}
\right)
+\mathcal{O}\left(\epsilon^{-1}\right)
\\&
\qFF{AF3} = 
-\frac{11}{3\epsilon^7}
+\frac{1}{\epsilon^6}\left(2 \zeta_2-\frac{265}{9}\right)
+\frac{1}{\epsilon^5}\left(
    \vphantom{\frac{1}{1}} 26 \zeta_3
    +6 \zeta_2
\right.\nonumber\\&\left.
    -\frac{18293}{108}
\right)
+\frac{1}{\epsilon^4}\left(
    \frac{78}{5} \zeta_2^2
    +\frac{3044}{9} \zeta_3
    -\frac{343}{6} \zeta_2
    -\frac{510383}{648}
\right)
\nonumber\\&
+\frac{1}{\epsilon^3}\left(
    182\zeta_5
    -96 \zeta_3 \zeta_2
    +\frac{10282}{45} \zeta_2^2
    +\frac{117305}{54} \zeta_3
\right.\nonumber\\&\left.
    -\frac{26125}{36} \zeta_2
    -\frac{12668057}{3888}
\right)
+\frac{1}{\epsilon^2}\left(
    \frac{13976}{315}\zeta_2^3
    -854 \zeta_3^2
\right.\nonumber\\&\left.
    +\frac{108332}{45}\zeta_5
    +\frac{724}{9} \zeta_3 \zeta_2
    +\frac{56921}{54} \zeta_2^2
    +\frac{3836057}{324} \zeta_3
\right.\nonumber\\&\left.
    -\frac{1107673}{216} \zeta_2
    -\frac{295187171}{23328}
\right)
+\mathcal{O}\left(\epsilon^{-1}\right)
\\&
\qFF{F4} = 
\frac{2}{3\epsilon^8}
+\frac{4}{\epsilon^7}
+\frac{1}{\epsilon^6}\bigg(
    \frac{59}{3}
    -\frac{4}{3} \zeta_2
\bigg)
+\frac{1}{\epsilon^5}\left(
    -\frac{272}{9} \zeta_3
\right.\nonumber\\&\left.
    +4\zeta_2
    +\frac{461}{6}
\right)
-\frac{1}{\epsilon^4}\left(
    \frac{296}{15}\zeta_2^2
    +\frac{508}{3}\zeta_3
    -\frac{164}{3}\zeta_2
    -\frac{6559}{24}
\right)
\nonumber\\&
+\frac{1}{\epsilon^3}\left(
    \frac{640}{9} \zeta_3 \zeta_2
    -\frac{3008}{15} \zeta_5
    -\frac{412}{5} \zeta_2^2
    -\frac{8390}{9} \zeta_3
    +\frac{1117}{3}\zeta_2
\right.\nonumber\\&\left.
    +\frac{1825}{2}
\right)
+\frac{1}{\epsilon^2}\bigg(
    \frac{19360}{27}\zeta _3^2
    -\frac{6784}{315}\zeta_2^3
    -\frac{8876}{5} \zeta_5
    -\frac{880}{3} \zeta_3 \zeta_2
\nonumber\\&
    -\frac{868}{3} \zeta_2^2
    -\frac{30826}{9} \zeta_3
    +\frac{7733}{4}\zeta_2
    +\frac{298691}{96}
\bigg)
+\mathcal{O}\left(\epsilon^{-1}\right)
\\&
\qFF{d4AF} = 
\frac{1}{\epsilon^2}\left(
    12 \zeta_3^2
    +\frac{248}{35} \zeta_2^3
    -\frac{110}{3}\zeta_5
    -\frac{4}{3} \zeta_3
    +4 \zeta_2
\right)
\nonumber\\&
+\mathcal{O}\left(\epsilon^{-1}\right)
\end{align}
for the non-singlet quark form factor,
\begin{align}
\label{eq:ffquarksinglet}
&\qFF{Ad3FF} = 
\frac{1}{\epsilon}\left(
    -\frac{7040}{3}\zeta_5
    -\frac{176}{5} \zeta_2^2
    +\frac{1232}{3} \zeta_3
    +880 \zeta_2
    +352
\right)
\nonumber\\&
+\mathcal{O}\left(\epsilon^{0}\right)
\\&
\qFF{Fd3FF} = 
\frac{1}{\epsilon^2}\left(
    \frac{1280}{3}\zeta_5
    +\frac{32}{5} \zeta_2^2
    -\frac{224}{3} \zeta_3
    -160 \zeta_2
    -64
\right)
\nonumber\\&
+\frac{1}{\epsilon}\left(
    \frac{5504}{3} \zeta_3^2
    +\frac{155648}{315} \zeta_2^3
    -\frac{6272}{9} \zeta_5
    -480 \zeta_3 \zeta_2
    +\frac{656}{3} \zeta_2^2
\right.\nonumber\\&\left.
    -\frac{2512}{9} \zeta_3
    -\frac{4880}{3} \zeta_2
    -\frac{3008}{3}
\right)
+\mathcal{O}\left(\epsilon^{0}\right)
\end{align}
for the singlet quark form factor, and
\begin{align}
\label{eq:ffgluon}
&\gFF{A3} = 
\frac{2}{3\epsilon^7}
-\frac{271}{162\epsilon^6}
-\frac{1}{\epsilon^5}\left(\frac{31}{27} \zeta_2+\frac{9329}{972}\right)
+\frac{1}{\epsilon^4}\left(
    \frac{6293}{972}
\right.\nonumber\\&\left.
    +\frac{1253}{81} \zeta_2
    -\frac{583}{27} \zeta_3
\right)
+\frac{1}{\epsilon^3}\left(
    -\frac{2198}{135} \zeta_2^2
    +\frac{11651}{243} \zeta_3
\right.\nonumber\\&\left.
    +\frac{1727}{81}\zeta_2
    +\frac{7863881}{34992}
\right)
+\frac{1}{\epsilon^2}\left(
    \frac{4570}{81} \zeta_3 \zeta_2
    -\frac{5213}{90} \zeta_5
\right.\nonumber\\&\left.
    +\frac{4982}{81} \zeta_2^2
    +\frac{126722}{729} \zeta_3
    -\frac{350995}{1458} \zeta_2
    +\frac{217942129}{209952}
\right)
\nonumber\\&
+\frac{1}{\epsilon}\left(
    \frac{121679}{162} \zeta_3^2
    +\frac{109234}{945} \zeta_2^3
    +\frac{209093}{1620} \zeta_5
    -\frac{185155}{486}\zeta_3 \zeta_2
\right.\nonumber\\&\left.
    +\frac{95137}{540} \zeta_2^2
    +\frac{778241}{2916} \zeta_3
    -\frac{24361195}{17496} \zeta_2
    -\frac{929929}{46656}
\right)
\nonumber\\&
+\mathcal{O}\left(\epsilon^0\right)
\\&
\gFF{A2F} = 
-\frac{22}{9\epsilon^5}
+\frac{1}{\epsilon^4}\bigg(\frac{176}{9}\zeta_3-\frac{733}{54}\bigg)
+\frac{1}{\epsilon^3}\bigg(
    \frac{64}{5} \zeta_2^2
    +\frac{157}{18} \zeta_2
\nonumber\\&\left.
    -\frac{1901}{27} \zeta_3
    +\frac{40735}{648}
\right)
-\frac{1}{\epsilon^2}\left(
    \frac{685}{9} \zeta_5
    +\frac{532}{9} \zeta_3 \zeta_2
    +\frac{763}{15} \zeta_2^2
\right.\nonumber\\&\left.
    +\frac{29530}{81} \zeta_3
    +\frac{4469}{108} \zeta_2
    -\frac{3167059}{3888}
\right)
+\frac{1}{\epsilon}\left(
    -\frac{2896}{21} \zeta_2^3
\right.\nonumber\\&\left.
    -\frac{18425}{27} \zeta_3^2
    -\frac{22331}{27} \zeta_5
    +\frac{7582}{27} \zeta_3\zeta_2
    -\frac{20827}{90} \zeta_2^2
\right.\nonumber\\&\left.
    -\frac{274931}{972} \zeta_3
    -\frac{375197}{648} \zeta_2
    +\frac{73947103}{23328}
\right)
+\mathcal{O}\left(\epsilon^0\right)
\\&
\gFF{AF2} = 
-\frac{11}{8\epsilon^3}
+\frac{1}{\epsilon^2}\left(330\zeta_5-\frac{407}{2}\zeta_3-\frac{1537}{24}\right)
\nonumber\\&
+\frac{1}{\epsilon}\left(
    554 \zeta_3^2
    +\frac{73096}{315} \zeta_2^3
    +360\zeta_5
    +32 \zeta_3 \zeta_2
    -\frac{1349}{10}\zeta_2^2
\right.\nonumber\\&\left.
    -\frac{3919}{3} \zeta_3
    +\frac{607}{12} \zeta_2
    -\frac{38669}{96}
\right)
+\mathcal{O}\left(\epsilon^0\right)
\\&
\gFF{F3} = 
\frac{69}{4\epsilon }
+\mathcal{O}\left(\epsilon^{0}\right)
\\&\gFF{d4AF} 
=
\frac{1}{\epsilon ^2}\left(
    \frac{40}{3}\zeta_5
    +\frac{8}{3} \zeta_3
    -8 \zeta_2
\right)
+\frac{1}{\epsilon}\left(
    \frac{1808}{315} \zeta_2^3
    -\frac{152}{3} \zeta_3^2
\right.\nonumber\\&\left.
    +\frac{3860}{9}\zeta_5
    -152 \zeta_3 \zeta_2
    -\frac{308}{15} \zeta_2^2
    -\frac{5312}{9} \zeta_3
    +8 \zeta_2
    +\frac{152}{3}
\right)
\nonumber\\&
+\mathcal{O}\left(\epsilon^{0}\right)
\\&
\gFF{A4} =
\frac{2}{3\epsilon^8}
-\frac{11}{3\epsilon^7}
+\frac{1}{\epsilon^6}\left(\frac{2}{3} \zeta_2+\frac{137}{648}\right)
+\frac{1}{\epsilon^5}\left(
    -\frac{38}{9} \zeta_3
\right.\nonumber\\&\left.
    +\frac{341}{54} \zeta_2
    +\frac{52775}{1944}
\right)
+\frac{1}{\epsilon ^4}\left(
    \frac{5}{18} \zeta_2^2
    +\frac{605}{54}\zeta_3
    -\frac{1889}{54} \zeta_2
\right.\nonumber\\&\left.
    +\frac{11383}{243}
\right)
+\frac{1}{\epsilon^3}\left(
    \frac{1082}{15}\zeta_5
    +\frac{23}{3} \zeta_3 \zeta_2
    +\frac{517}{27} \zeta_2^2
    +\frac{20405}{972}\zeta_3
\right.\nonumber\\&\left.   
    -\frac{114673}{972}\zeta_2
    -\frac{5347817}{17496}
\right)
+\frac{1}{\epsilon^2}\left(
    \frac{95198}{945} \zeta_2^3
    +\frac{5413}{27} \zeta_3^2
\right.\nonumber\\&\left.
    +\frac{1199}{81} \zeta_3 \zeta_2
    -\frac{64669}{60} \zeta_5
    -\frac{7493}{180} \zeta_2^2
    +\frac{653867}{729} \zeta_3
\right.\nonumber\\&\left.
    +\frac{175724}{729} \zeta_2
    -\frac{257277595}{104976}
\right)
+\mathcal{O}\left(\epsilon^{-1}\right)
\\&
\gFF{d4AA} = 
\frac{1}{\epsilon^2}\left(
    12 \zeta_3^2
    +\frac{248}{35}\zeta_2^3
    -\frac{110}{3}\zeta_5
    -\frac{4}{3} \zeta_3
    +4 \zeta_2
\right)
\nonumber\\&
+\mathcal{O}\left(\epsilon^{-1}\right)
\end{align}
 for the gluon form factor.
\bibliography{ff4lcusp}

\end{document}